\documentclass[letterpaper, 12pt]{article}%
\usepackage{graphicx}%
\usepackage[letterpaper, margin=1in]{geometry}
\usepackage{mathptmx}
\usepackage{mathtools}
\usepackage{amsmath}
\usepackage{amssymb}

\usepackage{subfig}
\usepackage{tikz} 
\usepackage{algorithm}
\usepackage[noend]{algpseudocode}

\usepackage{xcolor}

\usepackage{verbatim}
\usepackage{natbib}

\usepackage{mycommands}

\newtheorem{numpropinner}{Proposition}
\newenvironment{numprop}[1]{%
  \renewcommand\thenumpropinner{#1}%
  \numpropinner
}{\endnumpropinner}

\usepackage{setspace}
\usepackage[compact]{titlesec} 
\titlespacing{\section}{2pt}{2pt}{2pt} 
\usepackage{titling}
\title{A flexible and interpretable spatial covariance model for data on graphs}
\author{Michael F. Christensen and Peter D. Hoff}
\date{\small Department of Statistical Science, Duke University \\
Corresponding author: Michael Christensen, Email: michael.f.christensen@duke.edu \\
\today
} 

\DeclareMathAlphabet{\mathbfit}{OML}{cmm}{b}{it}
\DeclareMathOperator*{\argmin}{arg\,min} 

\begin{document}
\setstretch{1.7}
\maketitle
\vspace{-2em}

\begin{abstract}
    Spatial models for areal data are often constructed such that all pairs of adjacent regions are assumed to have near-identical spatial autocorrelation. In practice, data can exhibit dependence structures more complicated than can be represented under this assumption. In this article we develop a new model for spatially correlated data observed on graphs, which can flexibly represented many types of spatial dependence patterns while retaining aspects of the original graph geometry. Our method implies an embedding of the graph into Euclidean space wherein covariance can be modeled using traditional covariance functions, such as those from the Mat\'{e}rn family. We parameterize our model using a class of graph metrics compatible with such covariance functions, and which characterize distance in terms of network flow, a property useful for understanding proximity in many ecological settings. By estimating the parameters underlying these metrics, we recover the ``intrinsic distances" between graph nodes, which assist in the interpretation of the estimated covariance and allow us to better understand the relationship between the observed process and spatial domain. We compare our model to existing methods for spatially dependent graph data, primarily conditional autoregressive models and their variants, and illustrate advantages of our method over traditional approaches. We fit our model to bird abundance data for several species in North Carolina, and show how it provides insight into the interactions between species-specific spatial distributions and geography.
    

\end{abstract}

\section{Introduction}

Models for spatially indexed data traditionally assume that observations close to one another in space are more highly correlated than observations which are far apart. While models for point-indexed data within continuous space frequently characterize spatial correlation as a function of geographic distance between observations, models for area-indexed data, also known as areal data, tend to define spatial dependence as a function of the adjacency structure of the graphical representation of the spatial domain's partition \citep{verhoef2018a}. Such models often assume that the correlation between observations made at adjacent regions is the same throughout the network, an assumption that is often inappropriate for real world data \citep{guttorp1994}.

As an example of this we consider a simple exploratory data analysis made using data which was downloaded and processed from the eBird database. This database is a project managed by the Cornell Lab of Ornithology that crowd-sources bird observation data by allowing any user to record and submit their bird watching observations to the website. Observations are recorded in the form of user-submitted, location-indexed checklists containing the counts of each species observed by the bird watcher along with the time spent bird watching at each location \citep{ebird}. The data used here consist of the county-level number of combined observations of laughing gulls, a species common along the Atlantic coast, made in eastern North Carolina during each month between January 2018 and December 2020. During this time period, there were nearly 350,000 individual checklists submitted within the state of North Carolina, representing over 400,000 combined hours of bird watching. 

\begin{figure}[t!]
    \centering
    \includegraphics[width=15cm]{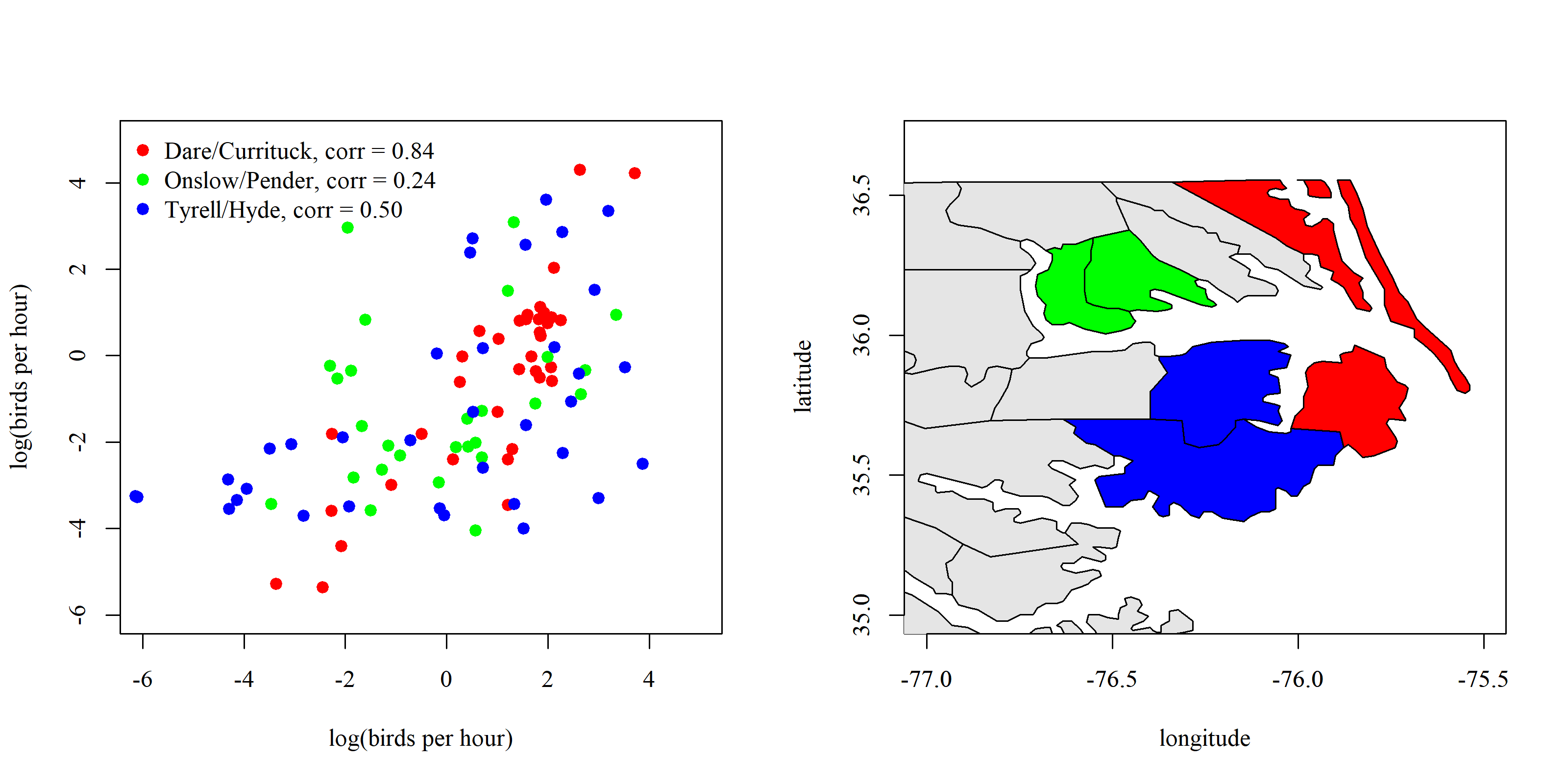}
    \caption{Three color-coded pairs of counties in North Carolina. Points correspond to the rate of laughing gull observation within a pair of counties during a month. Dare and Currituck counties (red) exhibit stronger correlation than Onslow and Pender counties (green) or Tyrell and Hyde counties (blue).}
    \label{ns_evid}
\end{figure}

To evaluate whether the degree of autocorrelation is consistent across our spatial domain we evaluate the month-to-month correlation in the data between pairs of adjacent counties. To account for the potential impact of environmental covariates on this simple analysis we restrict our attention the twenty coastal counties (as designated by the \citet{cama1974}) of North Carolina. Figure \ref{ns_evid} depicts the monthly log rate of observation for three pairs of adjacent counties, with each point corresponding to the frequency of observation at a pair of counties within a given month. If the assumption of constant spatial autocorrelation between adjacent counties were correct, then the correlations between adjacent counties should be approximately equal along the entirety of the coastline. As can be seen in the figure, the log rates of laughing gull observations in the adjacent Dare and Currituck counties are more correlated (0.84) than those of Onslow and Pender counties (0.24). To test this more more rigorously, we computed the between-county correlations for all 39 pairs of adjacent coastal counties. For each of those pairs, we obtained the statistic $z^\rho_i$ given by Fisher transformation of the Pearson correlation coefficient (known to be approximately normal \citep{fisher1921}) using the monthly log rates of observations at each county. An F-test of $H_0: \rho_1 = ... =\rho_{39}$ based on the statistics $\{z^\rho_i\}_{1:39}$ resulted in $p < 0.01$, indicating significant evidence of differences in the correlations between adjacent coastal counties, and implying that standard approaches to modeling areal data may fail to reflect this.

The majority of statistical methods designed to account for changing patterns of spatial dependence within a domain have been developed specifically for point referenced data. However, areal data, such as the county-level eBird data discussed in our example, are also common within many applications, either because some random variables are observed only at a regional level, or because raw point-referenced data might be aggregated and processed into areal or gridded regions out of convenience or due to scientific or policy interest in questions at the county, state, or national level. Such data are representable using graphs, with a node representing each region and edges connecting each pair of adjacent regions.

Conditional autoregressive (CAR) models, first introduced by \citet{besag1974}, are commonly used for modeling spatially dependent data observed on a graph, but are potentially limited when faced with more complex spatial dependence patterns. Given a random vector $\by$ with $y_1,...,y_p$ being observations at each of the $p$ nodes (or vertices) of some graph $G = (V,E)$, where $V$ contains the set of nodes, and $E$ is set of edges between those nodes, a CAR model specifies a Markov random field in which each element of $\by$ has a conditional distribution determined by the values at neighboring nodes, for example

\begin{equation}
    y_i|\by_{-i} \sim N\left( \sum_{j\sim i} \kappa w_{ij} y_j,\sigma^2\right)
\end{equation}
where $j\sim i$ indicates that node $j$ is adjacent to node $i$, $\bW$ is a $p \times p$ adjacency or weights matrix (where $w_{ij} > 0$ if $j\sim i$ else $w_{ij} = 0$), $\kappa$ controls the degree of spatial correlation (with $|\kappa| \leq 1$) and $\sigma^2 > 0$ is a variance parameter. This results in a marginal distribution for $\by$ of

\begin{equation}
    \by \sim N(\bzero, \sigma^2 (\bI_n - \kappa \bW)^{-1}).
\end{equation}

The CAR model may be thought of as the spatial generalization of the first order autoregressive (AR1) model often used in time series modeling. There are many varieties of CAR model that have been developed for use in generalized mixed models \citep{bym1991}, with multivariate data \citep{gelfand2003}, and which can utilize weighting schemes incorporating geographic and covariate information \citep{hanks2013, ejigu2020}. Most of the literature regarding spatially dependent graph data is limited to various forms of autoregressive models \citep{verhoef2018b}. The majority of CAR models and their variants tend to to treat all first-order neighbors within a graph as equally proximate due to the use of graph adjacency structure alone in their definition of spatial dependence. This will rarely result in effective representation of more complex spatial processes. In instances where more complex weighting schemes have been utilized, such as in \citet{hanks2013}, and \citet{ejigu2020}, edge weights are defined as linear functions of a small set of environmental covariates. While such approaches allow for more flexibility than those using adjacency alone, the space of possible weighting matrices under those models is restricted by the column space defined by the collection of environmental covariates being used. There are also instances in which covariates associated with graph edges may not be available to researchers, in which case a model must be defined using only the structure of the graph itself. One computational advantage of the CAR modeling framework is that the induced precision matrix will generally be sparse due to the assumption of independence for each observation conditional on the values of its neighbors. This assumption can come with trade-offs in terms of predictive efficiency; \citet{gramacy2015}, \citet{katzfuss2021}, \citet{guinness2018} all demonstrate the shortcomings of models which condition only on local or nearest neighbor structure (as is the case in CAR models) in comparison to those which also condition on more distant observations. For more discussion on CAR models see \citet{verhoef2018a} and \citet{lee2011} for review papers which provide detail on the use, development, and interpretation of these models and their variants. To address the concerns regarding limitations of CAR models for use with more complex spatial processes, we propose a model for the spatial covariance of graph data using a different set of assumptions, which we will describe in the following sections. We will also illustrate that our proposed method has greater capacity to represent certain correlation structures that may be of interest to statisticians and ecologists, especially in contrast to the most commonly used CAR variants.

Another approach to modeling the spatial random effects that may be present in areal data include basis function-based methods such as that of \citet{hughes2013}, though such approaches are generally structured more around the estimation of the spatial random effects themselves rather than the covariance structure that produced them. Additionally, such approaches often produce overly-smooth spatial random effects surfaces unless a very large number of basis functions are employed. 

Our goal within this article is not only to present a new covariance model for areal data that is flexible enough to account for varying patterns of spatial dependence across a domain, but to ensure that said model has interpretive value as well. The relationship between distance and spatial autocorrelation is often explicitly defined in the modeling of data observed within continuous space, but this relationship is complicated in the areal data setting, where the definition of a metric between regions is less straightforward \citep{cressie1993}. When considering our earlier example, it may be reasonable to imagine that from the perspective of a laughing gull within a coastal county, the changing environment as one moves inland from the coast may act as a barrier to movement, and that as a result, a neighboring coastal county is in some senses ``closer" than a neighboring inland county.
This proximity should be reflected in the data by a greater degree of spatial correlation between regions with smaller ``intrinsic distances." We could explicitly represent this relationship by parameterizing a unique distance metric characterizing the intrinsic distances between all counties within our spatial domain, and defining the covariance between counties as a simple function of this metric. This approach may be viewed as similar to the class of methods for modeling nonstationarity in continuous space in which spatial covariance is modeled as a stationary and isotropic function of a distance metric that arises from an estimated transformation of the geographic space in which the data were originally observed. See \citet{sampson1992,schmidt2003} and \citet{bornn2012} for examples of this approach.

Within this article we present a covariance model for graph data built under the idea that complex patterns of spatial dependence can be represented using a simple covariance function applied to a latent matrix of intrinsic distances between graph nodes. Fitting the model involves the estimation of a distance metric over our graph which in turn corresponds to an embedding of our graph into a high-dimensional Euclidean space. Covariance is then assumed to be a function of the distances between nodes in the embedding. In addition to being more flexible than most CAR model-based approaches, we find that intrinsic distance framework makes our model simple and intuitive to interpret. In the next section of this article we establish some of the necessary background information regarding covariance functions and graph metrics. In particular, we highlight a class of graph metrics which are compatible with most traditional covariance functions such as those from the Mat\'{e}rn family \citep{matern1960}. In Section 3 we formally introduce our model, and provide some detail on its interpretation as well as its implementation. In Section 4 we provide a case study using the eBird data discussed in the Introduction. We fit our model to bird abundance data for several species common to the state of North Carolina, and demonstrate the manner in which our model can provide valuable insight into the interaction between a spatial process and the underlying geography via the posterior distance matrix of the graph on which a process is observed. We conclude this article with a discussion of our model's strengths and limitations in comparison to existing methods, as well as potential extensions for future work.

\section{Background}
Suppose we are given a graph $G = (V,E)$ and random variables $\{Y_v: v \in V\}$ observed at the nodes $V$ of $G$. In this article we propose a model for $\Sigma_{vv'} = \text{Cov}(Y_v,Y_{v'})$ which is described visually in Figure \ref{overview}. The covariance matrix $\bSigma$ is parameterized by $\bW$, a matrix of unknown positive edge weights, and $\sigma^2 > 0$, a scale parameter. The parameter $\bW$ is an element of $\mathcal{W}_G$, the space of all possible edge weight matrices given $G$, and $\bW$ is mapped to a distance matrix $\bD$ via a graph metric $\Delta(\cdot)$. $\bD$ is in turn mapped to a symmetric positive definite matrix $\bSigma$ using a covariance function $C(\cdot)$ which also takes the parameter $\sigma^2 > 0$ as input. In the following two subsections, we establish the background information necessary to justify our model before presenting it formally in Section 3.

\begin{figure}[ht]
\centering
\begin{tikzpicture}[node distance={25mm}, ultra thick, main/.style = {draw, circle,minimum size = 2cm}] 
\node[] (2) at (2,0) {$\bW \in \mathcal{W}_G$};
\node[] (3) at (4,0) {$\bD$};
\node[] (4) at (6,0) {$\bSigma$};
\node[] (5) at (4.25,-.5) {$\sigma^2>0$};

\draw[ultra thick, ->] (2) -- node[midway, above, sloped, pos=.5] {\large $\Delta$} (3); 
\draw[ultra thick,->] (3) -- node[midway, above, sloped, pos=.5] {\large $C$} (4);
\draw[ultra thick] (4.9,-.4) -- (5.4,0);

\end{tikzpicture}
\caption{Visual representation of the model presented in this article. The edge weight parameter $\bW$ is mapped to a distance matrix, which in conjunction with the scale parameter $\sigma^2$ is mapped to the covariance matrix $\bSigma$. }\label{overview}
\end{figure}

\subsection{Covariance Functions and Distance}

Let $\mathcal{D}_p$ be the space of metric distance matrices between $p$ points. That is, if $\bD \in \mathcal{D}_p$ then $\bD$ is a $p \times p$ matrix where all diagonal elements of $\bD$ are zero, the off-diagonal entries are positive, $\bD$ is symmetric, and all elements of $\bD$ satisfy the triangle inequality (i.e. for any $\{i,j,k\} \subset \{1...p\} $ $d_{ij} \leq d_{ik} + d_{kj}$). In our model we define the $p \times p$ covariance matrix $\bSigma$ as $\bSigma = C(\bD),$ for $\bD \in \mathcal{D}_p$, where $C(\cdot)$ is a covariance function which maps distance matrices to $\mathcal{S}^+_p$, the space of $p \times p$ symmetric positive definite matrices. To allow for flexibility in the covariance, we will treat the distance matrix of a graph as an unknown parameter to be estimated from the data. This in contrast to most spatial models, which take the distances between observations as known and fixed. Instead we will parameterize distance as a function of edge weights on the graph, which are estimated from the data. This approach allows us define covariance structures for graph data that are considerably more complex than ones obtained using the neighborhood structure of the graph alone.

Covariance functions defined only by the distances between locations, and not the locations themselves, are said to be stationary and isotropic. These covariance functions are often defined such that they can be applied element-wise to a distance matrix. For $\bD \in \mathcal{D}_p$ we can also define $\bSigma$ by $\Sigma_{ij} = c(d_{ij})$ where $c(\cdot)$ is some function applied to a distance between two points. For $\bSigma$ to be a valid covariance matrix it must be symmetric and positive definite, and thus $c(\cdot)$ must be chosen carefully with respect to $\bD$. There are many well known covariance functions that can be used to produce positive definite matrices using distances as inputs. Functions from the Mat\'{e}rn covariance family, originally developed by \citet{matern1960} and given by
\begin{equation}\label{matern}
    c_\nu(d) = \sigma^2\frac{2^{1-\nu}}{\Gamma(\nu)}\left( \sqrt{2\nu}\frac{d}{\tau}\right)^\nu K_\nu \left( \sqrt{2\nu}\frac{d}{\tau}\right)
\end{equation}
are among the most commonly utilized within spatial statistics \citep{cressie1993,banerjee2003}. Here the input $d$ is a distance between two points, $\sigma^2$ is a scale parameter, $\tau$ is a range parameter, $\nu$ is a smoothness parameter, and $K_\nu$ is a modified Bessel function of the second kind of order $\nu$. The Mat\'{e}rn family includes the commonly used exponential and Gaussian (also known as the radial basis kernel or squared exponential) covariance functions \citep{banerjee2003}. If $c(\cdot)$ is a member of the Mat\'{e}rn family of covariance functions it is necessary for the distances contained in $\bD$ to be Euclidean in order to ensure that $\bSigma$ is a positive definite matrix \citep{verhoef2018c}. Here we define the space of Euclidean distance matrices $\mathcal{D}^E_p \subset \mathcal{D}_p$ such that if $\bD \in \mathcal{D}^E_p$, there exists $\{\bx_1, ..., \bx_p\} \subset \mathbb{R}^k$ for some $k$ such that $d_{ij} = \|x_i - x_j\|$.
In our model, for a given graph $G = (V,E)$ with $p$ nodes we will require that $\bD$, the matrix containing pairwise distances between the nodes of $G$, satisfies $\bD \in \mathcal{D}^E_p$ for some p. This also implies that there exists an embedding of $G$ into a Euclidean space such that the graph distances are equal to the distances between nodes in the embedding.

The fact that many of the most common covariance functions (including those from the Mat\'{e}rn family) require Euclidean distances as inputs to guarantee positive definiteness is reasonably well known, and can be confirmed when one considers Bochner's theorem, which provides a necessary and sufficient condition for functions on Euclidean vectors to be positive definite (see \citet{gihman1974}). However, this requirement has at times been disregarded by researchers who were interested in utilizing common spatial modeling tools with non-Euclidean distances, which can result in non-positive definite covariance matrices and negative prediction variances in applications. Such scenarios arise especially often in analyses of graph and network data, where the most commonly used metrics, such as the shortest path distance, often result in non-Euclidean distances. \citet{verhoef2018c} discusses this issue and provides several examples of published articles which failed to consider the non-compatibility of their chosen covariance functions and distance metrics.

\subsection{Graph Metrics}

To parameterize the covariance of random variables observed on a graph by applying the Mat\'{e}rn covariance function to a matrix of estimated distances, we must be careful that $\bD$ is parameterized in a manner that ensures that it is Euclidean over the entirety of the parameter space. To do this, we must consider what graph metrics produce Euclidean distances and how their properties will influence model interpretation.

Given an undirected, connected, simple graph $G = (V,E)$ with $p$ nodes, we define a class of distance matrices parameterized by the symmetric edge weights matrix $\bW$ of $G$. Many graph metrics are functions of $\bW$ where $w_{ij} > 0$ if node $i$ is adjacent to node $j$ and $w_{ij} = 0$ otherwise \citep{jungnickel2012}. We define $\mathcal{W}_G$ to be the space of all symmetric edge weights matrices possible given $G$, and note that all information regarding the edges of $G$ is contained in each $\bW \in \mathcal{W}_G$ based on whether or not an element of $\bW$ is equal to zero. The distance matrix $\bD$ for a graph $G$ is given by $\bD = \Delta(\bW)$ where $\Delta(\cdot)$ is the function for a particular metric which maps an edge weight matrix to a distance matrix. 

For a given graph $G$ and metric $\Delta(\cdot)$ we define $\mathcal{D}^\Delta_G$ as the space of distance matrices which are possible to obtain under different edge weight configurations for $G$, that is $\mathcal{D}^\Delta_G = \{\bD =\Delta(\bW): \bW \in \mathcal{W}_G\}$. We have established that for $\bSigma = C(\bD)$ to be a valid covariance matrix with $C(\cdot)$ a Mat\'{e}rn family covariance function, $\bD$ must be a Euclidean distance matrix. Thus, in choosing a graph metric $\Delta(\cdot)$, we must ensure that $\mathcal{D}^\Delta_G \subset \mathcal{D}^E_p$. In doing so, we can specify a parameterization of $\bSigma$ in terms of an unobserved edge weights matrix $\bW \in \mathcal{W}_G$ with the property that $\bSigma = C(\Delta(\bW)) \in \mathcal{S}^+_p$ for all $\bW \in \mathcal{W}_G$. This also implies that because $\bD = \Delta(\bW) \subset \mathcal{D}_p^E$, $\bW$ characterizes an embedding of $G$ into Euclidean space. 
In Section 3 we establish that for an appropriate choice of metric, our covariance model is identifiable, which implies that the embedding of $G$ implied by $\bW$ is unique up to isometry.

The most commonly used graph metric is the shortest path distance \citep{chebotarev2011} (denoted here as $\Delta^{SP}(\cdot)$) which defines the distance between nodes $i$ and $j$ as

\begin{equation}
    d_{ij} = \Delta^{SP}(\bW)_{ij} =  \min_{p_{ij}  \in P_{ij}}\left( \sum_{e_{lk} \in p_{ij}} w_{lk} \right)
\end{equation}
where $P_{ij}$ is the set of all possible paths between nodes $i$ and $j$, but can intuitively be described as the sum of the weights along the ``shortest" path connecting those nodes. However, this metric generally produces non-Euclidean distance matrices \citep{klein1998} and as such $\bSigma = C(\bD)$ is not guaranteed to be positive definite if $C(\cdot)$ is Mat\'{e}rn.

A commonly used alternative graph metric is the resistance distance \citep{chebotarev2011}, denoted $\Delta^{\Omega}(\cdot)$, where the distance between nodes $i$ and $j$ is defined as

\begin{equation}
    d_{ij} = \Delta^\Omega(\bW)_{ij} = (\be_i - \be_j)^\top\bL^+(\be_i-\be_j)
\end{equation}
where $\bL^+$ is the Moore-Penrose generalized inverse of the graph Laplacian, defined as $\bL = \text{diag}(\bW\bone_p) - \bW$ where $\text{diag}(\cdot)$ is a function taking a $p$-length vector and returning a $p \times p$ diagonal matrix, $\bone_p$ is a vector of ones, and $\{\be_i\}_{1:p}$ are the standard basis vectors. This metric was introduced to the mathematical literature by \citet{klein1993} with roots in electrical physics and is based on the formula for calculating the effective resistance between nodes in a network of resistors.

Attractive properties of resistance distance include the fact that the distance between two nodes is reduced by the number of paths connecting said nodes, a property that shortest path distance does not possess \citep{klein1993}, and that it is proportional to commute time distance, i.e. the expected hitting time of a random walk between two nodes \citep{chandra1996}. Use of resistance distance is widespread within ecological settings as both of the above properties characterize a type of flow-based graph connectivity which corresponds nicely with existing models for the movement of individual organisms as well as animal populations \citep{thiele2018, peterson2019}.

Because $\bL^+$ is always a positive semi-definite matrix, resistance distance may be viewed as a squared Euclidean (or Mahalanobis) distance. \citet{klein1998} highlight that a related class of metrics of the form

\begin{equation}
    d_{ij} = \Delta^{(m)}(\bW)_{ij} = \sqrt{(\be_i - \be_j)^\top\{\bL^+\}^m(\be_i-\be_j)}
\end{equation}
where $m>0$ is an exponent (i.e  $\{\bL^+\}^2 = \bL^+\bL^+$), are guaranteed to produce Euclidean distance matrices for any $m>0$ and all $\bW \in \mathcal{W}_G$. Thus, all metrics of this form can be used in conjunction with any covariance function which requires Euclidean distances as inputs, such as those from the Mat\'{e}rn class. The square root of resistance distance is one of these metrics (when $m=1$), but we are particularly interested in using the second metric of this form (when $m=2$)\textemdash often referred to as the quasi-Euclidean metric\textemdash because it shares many of the same properties regarding graph connectivity with resistance distance and has been demonstrated to be well behaved in comparison to other metrics, especially when used with planar graphs which are typically seen in spatial and ecological settings \citep{zhu1996, ivanciuc2001}. The quasi-Euclidean metric also scales linearly with the inverse of the edge weights, that is, $c\bD = \Delta^{(2)}(\bW/c)$ for $c > 0$. This property is convenient for computation and enables greater parameter interpretability. Subsection 3.2 contains a numeric illustration of how the quasi-Euclidean metric operates within our model.

\section{Method}
\subsection{A Flexible Graph Covariance Model}

Given an undirected, connected, simple graph $G = (V,E)$ with $p$ nodes, we make repeated observations of the collection of random variables $\{Y_v: v \in V\}$; each repetition is recorded as $\by_i \in \mathbb{R}^p$ for $i \in 1,...,n$. These observations may be stored in the $n \times p$ data matrix $\bY$, where $y_{ij}$ is the $i$th observation at the $j$th node of $G$. We define $\bSigma$, the between-node covariance of each $\by_i$, in terms of the intrinsic distances between the nodes of $G$ which are in turn a function of the known structure of the graph $(V,E)$ and unobserved edge weight parameter matrix $\bW$, which is estimated from the data. We highlight that we consider the structure of the graph $G$ to be fixed. This stands in contrast to methods such as those presented in \citet{white2009} and \citet{ma2010} which treat the adjacency structure itself as something unknown to be estimated. Our model instead estimates the unknown edge weights for a fixed graph, though in some senses our approach has the ability to omit edges from the model by estimating their weights to be near zero. This choice may be particularly justified in settings involving animal and plant populations, where interaction or movement between non-physically adjacent regions can only occur through intermediate regions. Our parameterization of $\bSigma$ could be used in conjunction with any number of distributional assumptions for $\bY$, including temporal dependence or non-normality, but to make presentation concrete we initially define our model using the assumption that $\{\by_i\}_{1:n}$ are independent draws from a $N(\bzero_p,\bSigma)$ distribution. We will discuss possible extensions later in the article. We parameterize $\bSigma$ as follows:

\begin{equation}\label{model}
\begin{aligned}
    \Sigma_{jk} &= \sigma^2 \rho_\nu(d_{jk}) \\
    d_{jk} &= \sqrt{(\be_j - \be_k)^\top\{\bL^+\}^2(\be_j-\be_k)} \\
    \bL &= \text{diag}(\bW \bone_p) - \bW \\
    \sigma^2 &> 0, \; w_{jk} > 0  \text{ if } j \sim k, \text{ else } w_{jk} = 0, \text{ and } w_{jk} = w_{kj}.
\end{aligned}
\end{equation}
Under this model, the data $\bY$ comes from a mean-zero matrix-normal distribution with column covariance $\bSigma$ and row covariance $\bI_n$. The parameter $\bW$ is a $p \times p$ symmetric edge weight matrix, $\sigma^2$ is a scale parameter, and $\rho_\nu(\cdot)$ is the Mat\'{e}rn correlation function given below: 

\begin{equation}\label{our_mat}
    \rho_\nu(d) = \frac{2^{1-\nu}}{\Gamma(\nu)}\left( \sqrt{2\nu}d\right)^\nu K_\nu \left( \sqrt{2\nu}d\right).
\end{equation}
Common practice is to take the smoothness parameter $\nu$ as fixed rather than as a parameter to be estimated. In this article we set $\nu = 3/2$, which implies a spatial process that is smoother than those given by an exponential covariance function, but less smooth than those defined with Gaussian covariance; however, any value of $\nu > 0$ would be valid in this model \citep{stein1999}. 

Note that the correlation function defined above does not include an explicit range or spatial-decay parameter. Equation \ref{matern}, the standard form of the Mat\'{e}rn covariance function, contains the parameter $\tau$ whereas the correlation function we are using, given in Equation \ref{our_mat}, does not. Despite the lack of a parameter $\tau$, which directly scales the distances used as inputs in the correlation function, our parameterization of $\bSigma$ implicitly controls for the rate of spatial decay thru the edge weight matrix $\bW$: Let $\bD$ be the $p \times p$ distance matrix produced by applying the quasi-Euclidean metric to the edge weight matrix $\bW$. The third and fourth lines of Equation \ref{model} show how $\bD$ is obtained: let $\Delta^{(2)}(\cdot)$ be the function such that $\bD = \Delta^{(2)}(\bW)$. Recall that $c\bD = \Delta^{(2)}(\bW/c)$ for any positive constant $c$. The distances in our model are directly scaled by the magnitude of the edge weights, which are themselves unrestricted above zero rendering an additional parameter $\tau$ where correlation is a function of $d/\tau$ unnecessary. To interpret our model's estimate of the spatial decay in the covariance of $\bY$, we recommend re-scaling the distance matrix $\bD$ after fitting. If we define $\bD^s = \bD/\text{max}(\bD)$ and $\tau^s = 1/\text{max}(\bD)$, then the covariance defined by $\Sigma_{jk} = \rho_\nu(d^s_{jk}/\tau^s)$ is equal to the one defined $\Sigma_{jk} = \rho_\nu(d_{jk})$, and $\tau^s$ has equivalent interpretation to a range parameter in a covariance model for point-referenced data in which the geographic distances between observations were scaled to have a maximum distance of one. 

As parameterized, our model is identifiable:

\begin{numprop}{1}[Identifiability]
Using the parameterization for $\bSigma$ given in Equation \ref{model}, $(\sigma^2_1, \bW_1) \neq  (\sigma^2_2, \bW_2)$ implies $\bSigma_1 \neq \bSigma_2$ for all $\sigma^2_1,\sigma^2_2 > 0$ and all $\bW_1, \bW_2 \in \mathcal{W}_G$.
\end{numprop}
A proof of the model's identifiability is provided in the supplemental material for this article.

\subsection{Interpretation}

Rather than characterizing spatial covariance via a complex covariance function or kernel, our model uses a relatively simple covariance function, and pushes most of the modeling complexity into the estimation of latent distances between graph nodes, which are themselves functions of a set of edge weight parameters; nodes that are close to one another are more correlated, while nodes that are far apart are less correlated. The metric we utilize to define distances over the graph was chosen to ensure that the resulting covariance matrix will be positive definite for any combination of edge weight parameters, but what of the edge weight parameters themselves? Generally speaking, larger edge weight parameters indicate greater correlation between nodes in the regions of the graph where that edge is found, while smaller edge weight parameters indicate lower correlation in those regions. When edge weights approach zero, the corresponding edges are effectively omitted from the graph, which may be thought of as a type of automatic model selection for the design of the graph itself. 

\begin{figure}
\centering

\subfloat[]{
\begin{tikzpicture}[node distance={17mm}, thick, main/.style = {draw, circle}] 
\node[] at (0,0) {$G_1=$};
\node[] at (8,0) {$\longrightarrow \bSigma_1 = \begin{bmatrix}
1.00 & 0.43 & 0.51 & 0.43 & 0.31 \\ 
0.43 & 1.00 & 0.51 & 0.31 & 0.43 \\ 
0.51 & 0.51 & 1.00 & 0.51 & 0.51 \\ 
0.43 & 0.31 & 0.51 & 1.00 & 0.43 \\ 
0.31 & 0.43 & 0.51 & 0.43 & 1.00 \end{bmatrix}$};

\node[main] (3) at (2.5,0) {3}; 
\node[main] (1) [above left of=3] {1}; 
\node[main] (2) [above right of=3] {2}; 
\node[main] (4) [below left of=3] {4}; 
\node[main] (5) [below right of=3] {5}; 
\draw (1) -- node[midway, above, sloped, pos=.5] {0.33} (2); 
\draw (1) -- node[midway, above, sloped, pos=.5] {0.33} (3);
\draw (4) -- node[midway, above, sloped, pos=.5] {0.33} (1); 
\draw (2) -- node[midway, above, sloped, pos=.5] {0.33} (3); 
\draw (3) -- node[midway, above, sloped, pos=.5] {0.33} (4);
\draw (2) -- node[midway, above, sloped, pos=.5] {0.33} (5); 
\draw (5) -- node[midway, above, sloped, pos=.5] {0.33} (3);
\draw (5) -- node[midway, below, sloped, pos=.5] {0.33} (4); 
\end{tikzpicture}\label{ex_a}}

\subfloat[]{
\begin{tikzpicture}[node distance={17mm}, thick, main/.style = {draw, circle}] 
\node[] at (0,0) {$G_2=$};
\node[] at (8,0) {$\longrightarrow \bSigma_2 = \begin{bmatrix}
1.00 & 0.89 & 0.91 & 0.89 & 0.84 \\ 
  0.89 & 1.00 & 0.91 & 0.84 & 0.89 \\ 
  0.91 & 0.91 & 1.00 & 0.91 & 0.91 \\ 
  0.89 & 0.84 & 0.91 & 1.00 & 0.89 \\ 
  0.84 & 0.89 & 0.91 & 0.89 & 1.00   \end{bmatrix}$};

\node[main] (3) at (2.5,0) {3}; 
\node[main] (1) [above left of=3] {1}; 
\node[main] (2) [above right of=3] {2}; 
\node[main] (4) [below left of=3] {4}; 
\node[main] (5) [below right of=3] {5}; 
\draw (1) -- node[midway, above, sloped, pos=.5] {1.0} (2); 
\draw (1) -- node[midway, above, sloped, pos=.5] {1.0} (3);
\draw (4) -- node[midway, above, sloped, pos=.5] {1.0} (1); 
\draw (2) -- node[midway, above, sloped, pos=.5] {1.0} (3); 
\draw (3) -- node[midway, above, sloped, pos=.5] {1.0} (4);
\draw (2) -- node[midway, above, sloped, pos=.5] {1.0} (5); 
\draw (5) -- node[midway, above, sloped, pos=.5] {1.0} (3);
\draw (5) -- node[midway, below, sloped, pos=.5] {1.0} (4); 
\end{tikzpicture}\label{ex_b}}

\subfloat[]{
\begin{tikzpicture}[node distance={17mm}, thick, main/.style = {draw, circle}] 
\node[] at (0,0) {$G_3=$};
\node[] at (8,0) {$\longrightarrow \bSigma_3 = \begin{bmatrix}
1.00 & 0.90 & 0.41 & 0.90 & 0.24 \\ 
   0.90 & 1.00 & 0.41 & 0.78 & 0.25 \\ 
   0.41 & 0.41 & 1.00 & 0.41 & 0.24 \\ 
   0.90 & 0.78 & 0.41 & 1.00 & 0.25 \\ 
   0.24 & 0.25 & 0.24 & 0.25 & 1.00   \end{bmatrix}$};

\node[main] (3) at (2.5,0) {3}; 
\node[main] (1) [above left of=3] {1}; 
\node[main] (2) [above right of=3] {2}; 
\node[main] (4) [below left of=3] {4}; 
\node[main] (5) [below right of=3] {5}; 
\draw (1) -- node[midway, above, sloped, pos=.5] {2.0} (2); 
\draw (1) -- node[midway, above, sloped, pos=.5] {0.2} (3);
\draw (4) -- node[midway, above, sloped, pos=.5] {2.0} (1); 
\draw (2) -- node[midway, above, sloped, pos=.5] {0.2} (3); 
\draw (3) -- node[midway, above, sloped, pos=.5] {0.2} (4);
\draw (2) -- node[midway, above, sloped, pos=.5] {0.2} (5); 
\draw (5) -- node[midway, above, sloped, pos=.5] {0.2} (3);
\draw (5) -- node[midway, below, sloped, pos=.5] {0.2} (4); 
\end{tikzpicture}\label{ex_c}}

\subfloat[]{
\begin{tikzpicture}[node distance={17mm}, thick, main/.style = {draw, circle}] 
\node[] at (0,0) {$G_4=$};
\node[] at (8,0) {$\longrightarrow \bSigma_4 = \begin{bmatrix}
1.00 & 0.52 & 0.52 & 0.52 & 0.56 \\ 
   0.52 & 1.00 & 0.32 & 0.32 & 0.52 \\ 
   0.52 & 0.32 & 1.00 & 0.32 & 0.52 \\ 
   0.52 & 0.32 & 0.32 & 1.00 & 0.52 \\ 
   0.56 & 0.52 & 0.52 & 0.52 & 1.00  \end{bmatrix}$};

\node[main] (3) at (2.5,0) {3}; 
\node[main] (1) [above left of=3] {1}; 
\node[main] (2) [above right of=3] {2}; 
\node[main] (4) [below left of=3] {4}; 
\node[main] (5) [below right of=3] {5}; 
\draw (1) -- node[midway, above, sloped, pos=.5] {0.5} (2); 
\draw (1) -- node[midway, above, sloped, pos=.5] {0.5} (3);
\draw (4) -- node[midway, above, sloped, pos=.5] {0.5} (1); 
\draw (2) -- node[midway, above, sloped, pos=.5] {0.005} (3); 
\draw (3) -- node[midway, above, sloped, pos=.5] {0.005} (4);
\draw (2) -- node[midway, above, sloped, pos=.5] {0.5} (5); 
\draw (5) -- node[midway, above, sloped, pos=.5] {0.5} (3);
\draw (5) -- node[midway, below, sloped, pos=.5] {0.5} (4); 
\end{tikzpicture}\label{ex_d}}
\caption{The covariance matrices for a simple graph resulting from several different edge weight configurations when using a Mat\'{e}rn covariance function with $\nu = 3/2$.} \label{example}
\end{figure}
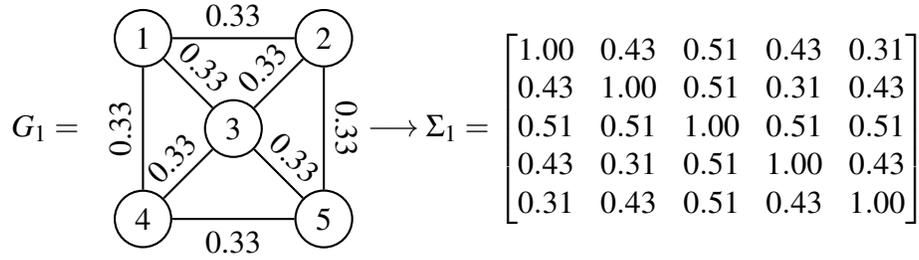
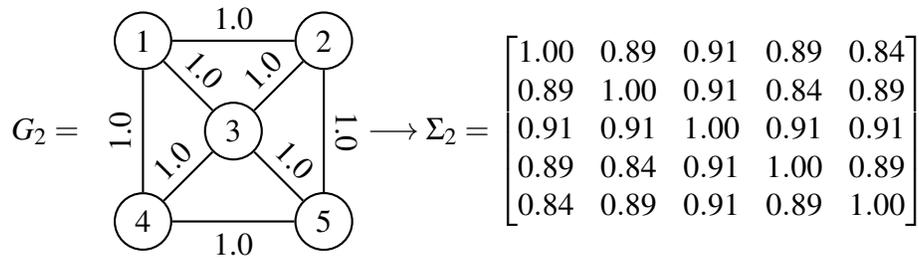
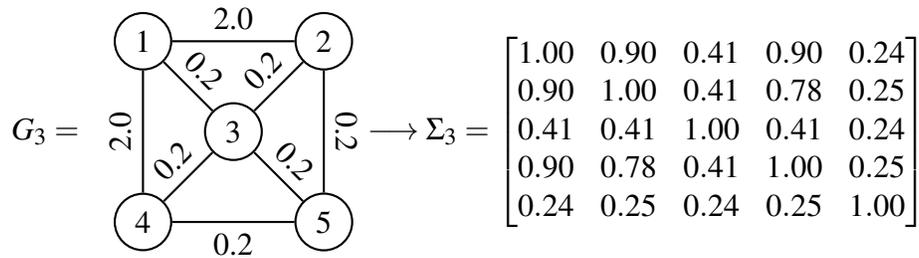
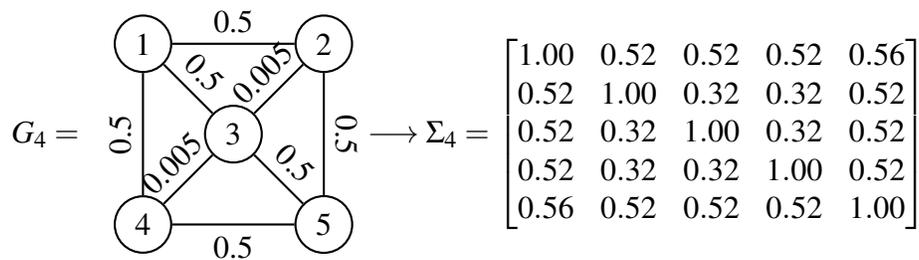

Figure \ref{example} contains an illustration of how edge weights impact covariance in our model. It depicts four sets of edge weights for the same five-node graph and shows the resulting correlation matrix when a Mat\'{e}rn correlation function with $\nu = 3/2$ is applied to the distance matrix produced by the edge weights. Subfigures \ref{ex_a} and \ref{ex_b} highlight how the scale of edge weights matter, with larger edge weights producing greater between-node correlation. Subfigure \ref{ex_c} demonstrates the potential flexibility of our model when certain edge weights are larger than others, with nodes along the path from 4 to 1 to 2 exhibiting greater intercorrelation than the rest of the graph. Subfigure \ref{ex_d} shows that it is possible under our model to obtain covariance structures where non-adjacent nodes (1 and 5) are more correlated (in effect closer together) than any other pair of first-order neighbors, which is possible due to the properties of the quasi-Euclidean metric, for which the distance between non-adjacent nodes is decreased with an increase in the number of sufficiently weighted connecting paths.

\subsection{Comparison to CAR models}
The covariance model specified here is quite flexible, especially in contrast to the commonly used graph-based methods which treat all first-order neighbors as equally proximate, such as most CAR models. We illustrate this with a simple example. The correlation component of our model for $\bSigma$ is determined fully by $\bW \in \mathcal{W}_G$, while $\sigma^2$ simply acts to scale the correlation matrix defined by $\bW$. Applying the quasi-Euclidean metric to $\bW$ in turn characterizes $\mathcal{D}^\Delta_G,$ the space of distance matrices possible under the model, which in turn can be used to define $\mathcal{S}^{(C \circ \Delta)}_G \subset \mathcal{S}^+_p$, the subset of covariance matrices that can be obtained under our model for a given graph, metric, and covariance function. Fitting our model to data results in us finding the ``best" $\hat{\bSigma}$ in $\mathcal{S}^{(C \circ \Delta)}$ to approximate $\bSigma$ as defined by maximizing likelihood, minimizing Bayesian loss, or some other approach. Just as with our model, other specifications of $\bSigma$ using fewer than $p(p-1)/2$ parameters (the number of free elements in a $p \times p$ covariance matrix)  characterize a proper subset of $\mathcal{S}^+_p$ containing all covariance matrices that are possible to obtain under that model. 

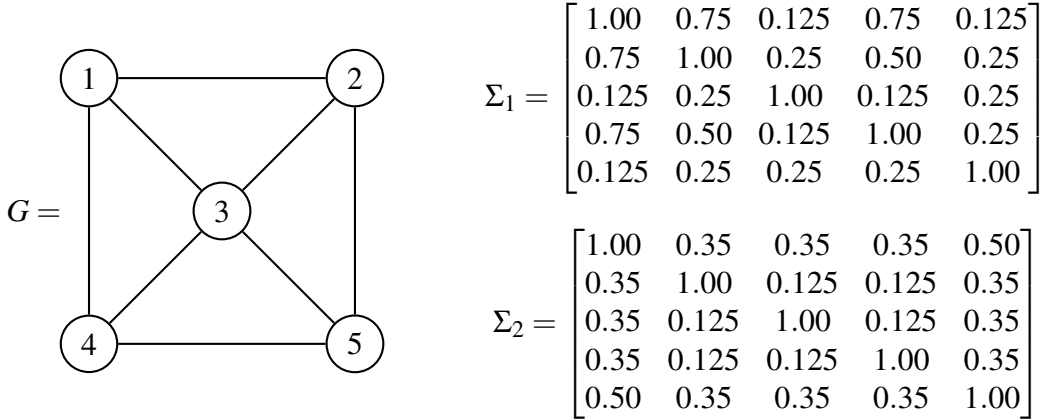
\begin{figure}
\centering

\begin{tikzpicture}[node distance={25mm}, thick, main/.style = {draw, circle}] 
\node[] at (-.5,0) {$G=$};
\node[] at (9.25,1.5) {$\bSigma_1 = \begin{bmatrix}
1.00 & 0.75 & 0.125 & 0.75 & 0.125 \\ 
0.75 & 1.00 & 0.25 & 0.50 & 0.25 \\ 
0.125 & 0.25 & 1.00 & 0.125 & 0.25 \\ 
0.75 & 0.50 & 0.125 & 1.00 & 0.25 \\ 
0.125 & 0.25 & 0.25 & 0.25 & 1.00 \end{bmatrix}$};
\node[] at (9.25,-1.5){$\bSigma_2 = \begin{bmatrix}
1.00 & 0.35 & 0.35 & 0.35 & 0.50 \\ 
0.35 & 1.00 & 0.125 & 0.125 & 0.35 \\ 
0.35 & 0.125 & 1.00 & 0.125 & 0.35 \\ 
0.35 & 0.125 & 0.125 & 1.00 & 0.35 \\ 
0.50 & 0.35 & 0.35 & 0.35 & 1.00 \end{bmatrix}$};

\node[main] (3) at (2,0) {3}; 
\node[main] (1) [above left of=3] {1}; 
\node[main] (2) [above right of=3] {2}; 
\node[main] (4) [below left of=3] {4}; 
\node[main] (5) [below right of=3] {5}; 
\draw (1) -- node[midway, above, sloped, pos=.5] {} (2); 
\draw (1) -- node[midway, above, sloped, pos=.5] {} (3);
\draw (4) -- node[midway, above, sloped, pos=.5] {} (1); 
\draw (2) -- node[midway, above, sloped, pos=.5] {} (3); 
\draw (3) -- node[midway, above, sloped, pos=.5] {} (4);
\draw (2) -- node[midway, above, sloped, pos=.5] {} (5); 
\draw (5) -- node[midway, above, sloped, pos=.5] {} (3);
\draw (5) -- node[midway, below, sloped, pos=.5] {} (4); 
\end{tikzpicture}

\caption{A graph $G$ and two correlation matrices used for model fit comparisons. We identify the closest approximations of $\bSigma_1$ and $\bSigma_2$ from the subsets of correlation matrices defined by multiple models.} \label{illustration}
\end{figure}

Figure \ref{illustration} depicts a graph $G$ with five nodes, and two correlation matrices, $\bSigma_1$ and $\bSigma_2$, which represent possible dependence structures for data observed on $G$. We note that neither $\bSigma_1$ or $\bSigma_2$ are elements of $\mathcal{S}^{(C \circ \Delta)}_G$. The matrix $\bSigma_1$ depicts a correlation structure in which one side of $G$ (the path connecting nodes 4-1-2) exhibits greater correlation than the rest of the graph, while $\bSigma_2$ depicts a correlation structure in which non-adjacent nodes (1 and 5) are more correlated with one another than any pair of adjacent nodes in $G$. These two configurations of $\bSigma$ were chosen to illustrate complex covariance structures which could be present within real world data, but are not meant to represent the full range of possible cases. In this example, we assume a multivariate normal, mean-zero distribution and find that best approximations of $\bSigma_1$ and $\bSigma_2$ from $\mathcal{S}^{(C \circ \Delta)}_G$ minimize Kullback-Leibler (KL) divergence to a greater degree than the equivalent best approximations from the CAR models we considered. Specifically, $\hat{\bSigma}_1$ and $\hat{\bSigma}_2$ are obtained by identifying the optimal set of edge weights such that

\begin{equation}
    \hat{\bSigma}_i = \argmin_{\bSigma} \text{KL}(N(\bzero,\bSigma),N(\bzero,\bSigma_i)), \; \bSigma \in \mathcal{S}^{(C \circ \Delta)}_G, \; i = 1,2
\end{equation}

The most widely used CAR variant is the first-order CAR model, which we will refer to as the CAR1 model; it incorporates only the basic adjacency structure of $G$ into its design and has a marginal distribution which defines
\begin{equation}\label{CAR1}
    \bSigma = \sigma^2(\bI_n-\kappa\bA)^{-1}, \quad \sigma^2 > 0, \; \kappa \in (-1,1)
\end{equation} 
where $\bA$ is the fixed adjacency matrix of $G$. Let $\mathcal{S}^{\text{CAR1}}_G$ be the space of covariance matrices possible to obtain under the CAR1 model. Given that the CAR1 model has only two parameters, it may be unfair to compare it to a more complex model such as the one proposed in this article, which has number of parameters equal to the number of edges in $G$ (8 in the case of $G$ from Figure \ref{illustration}) plus one. We thus propose a second more complex CAR model, which we refer to as the weighted CAR or CARw model, which represents the maximally flexible correlation structure possible under the CAR framework, (as characterized by this article) in which we treat all non-zero elements of the weights matrix as parameters, defining 
\begin{equation}\label{CARw}
    \bSigma = \sigma^2(\text{diag}(\bW\bone) - \kappa \bW)^{-1}, \quad \sigma^2 >0, \; \kappa \in (-1,1), \; \bW \in \mathcal{W}_G.
\end{equation}
Let $\mathcal{S}^{\text{CARw}}_G$ be the space of covariance matrices possible to obtain under this CAR model, which is specified using a weights matrix $\bW$ for which edge weights are estimated individually, just as with the model described in Section 3. This type of autoregressive model is uncommon within the literature, though a similar specification appears in \citet{smith2015}. In general, the limited number of existing CAR approaches utilizing a flexible weights matrix tend to define the elements of $\bW$ as some function of environmental covariates, geographic distances, or the sizes of areal units rather than estimating individual edge weights directly \citep{hanks2013, wang2016, verhoef2018a}.

For each of the three models discussed above and the two correlation matrices from Figure \ref{illustration}, we approximated the correlation matrix from each model which minimizes the KL-divergence between the distributions characterized by the true $\bSigma$ and the model estimate $\hat{\bSigma}$. This was done by generating 100,000 samples from $N_5(\bzero_5,\bSigma)$ and obtaining a Bayes estimator for $\bSigma$ using the simulated data (the details for obtaining this estimator are described in the following subsection). The Bayes estimator has been shown to concentrate about the parameter value which minimizes KL-divergence between a model and the true distribution as sample size increases \citep{zellner1988, gelman2014}. For this exercise we scaled each draw of $\bSigma$ within our sampler to ensure that it was a correlation matrix; this was done to improve the fairness of the comparison between our model and the CAR models which will generally have heterogeneous diagonal variance if specified as in equations \ref{CAR1} and \ref{CARw}.

Table \ref{ill_res} contains the KL-divergences between each model and the truth for $\bSigma_1$ and $\bSigma_2$. $\hat{\bSigma}$ is the approximate optimal parameter under each model in terms of KL-divergence. Lower KL-divergence indicates that the subset of covariance matrices defined by a particular model gets closer to the true data-generating distribution. As can be seen, our model is able to more closely approximate both $\bSigma_1$ and $\bSigma_2$ than the CAR variants we considered. The CAR1 model performs notably worse than our method, which is unsurprising given that it uses far fewer parameters and the KL-divergence does not account for model complexity. The CARw model has one more parameter than our model, yet still performs worse for both $\bSigma_1$ and $\bSigma_2$ with the gap in performance being more apparent for $\bSigma_2$. This is due in part to the fact that our model has greater flexibility than most autoregressive models to produce covariance matrices for which non-adjacent nodes are more correlated with one another than to any of the intermediate nodes due how the quasi-Euclidean metric defines distance in the presence of multiple viable paths between locations.

\begin{table}[t!]
\centering
\begin{tabular}{rrr}
  \hline
 &$\bSigma_1$ & $\bSigma_2$  \\ 
  \hline
  $\hat{\bSigma}^{(C \circ \Delta)}$ & \textbf{0.093} & \textbf{0.028} \\ 
  $\hat{\bSigma}^{\text{CAR1}}$ & 1.135 & 0.249  \\
    $\hat{\bSigma}^{\text{CARw}}$ & 0.150 & 0.111  \\ 

\end{tabular}
\caption{$\text{KL}(N(\bzero,\hat{\bSigma_i}),N(\bzero,\bSigma_i))$ for each model and correlation matrix. The KL-divergence of the best performing model for each ``true" correlation matrix is in bold.}\label{ill_res}
\end{table}

\subsection{Priors and Model Fitting}
While one could obtain parameter estimates using maximum likelihood estimatation, the complexity of the relationship between edge weights and the covariance matrix makes maximizing the likelihood challenging in comparison to Bayesian estimation using a Markov chain Monte Carlo (MCMC) approach. We suggest priors in accordance with the parameter space of our model which requires that $\sigma^2$ and all edge weight parameters $\{w_{jk}: j \sim k\}$ must be positive real numbers. If the nodes of the graph correspond to a known orientation in physical space (as is frequently the case when dealing with areal data) one can obtain the geographic distance matrix $\bD^{Geo}$ by calculating the physical distances between centroids of areal units. It is then possible to encode this geographic information into the prior for the edge weights, which may be helpful for model fitting, especially when dealing with a limited number of observations of the process over the graph. In such cases, we propose using the following priors:

\begin{equation}\label{prior}
\begin{aligned}
    \sigma^2 &\sim \text{Inverse-gamma}(a_{\sigma^2},b_{\sigma^2}) \\
    w_{jk} &\sim \text{Gamma}(a_w/d^{Geo}_{jk},b_w) \text{ if } j \sim k,\text{ else } w_{jk} = 0.
\end{aligned}
\end{equation}
Here $d^{Geo}_{jk}$ is the geographic distance between nodes $j$ and $k$, and the prior expectation of the associated edge weight is $a_w/(d^{Geo}_{jk}b_w)$ which is smaller for adjacent nodes which are far apart than for adjacent nodes which are closer in physical space. In instances where no geographic information is available about the graph on which data are observed, it is appropriate to simply use i.i.d. gamma priors for the edge weights. The choice of inverse-gamma prior for $\sigma^2$ is convenient for normally distributed $\bY$ due to built-in conjugacy. The choice to use independent gamma priors for all edge weights is somewhat simplistic, as it may be reasonable to assume some level of dependence between the weights of incident edges. Details for the MCMC sampler used to obtain posterior samples for $\sigma^2$ and edge weight matrix $\bW$ when fitting this model are provided in the supplemental material for this article. 

\subsection{Simulation Study}

To assess the practical capacity of our model to reasonably perform parameter estimation we conducted the following simulation study. We defined 5-by-5, 10-by-10, and 15-by-15 lattices
each having 40, 180, and 420 individual edges respectively. An edge weights matrix $\bW$ was randomly generated with each nonzero edge having a Gamma(3,3) distribution, and $n = 10$, 25, or 50 samples $\{\by_i\}_{1:n}$ were independently drawn from a normal distribution with mean zero and covariance $\bSigma(\sigma^2,\bW) + \psi \bI_p$, with $\psi$ acting as a nugget parameter, and with the setting $(\sigma^2 = 0.8, \psi = 0.2)$ representing a scenario in which the spatial signal to noise ratio is high and the setting $(\sigma^2 = 0.2, \psi = 0.8)$ a scenario in which the ratio is high.

All possible combinations of graph-size, samples generated, and signal-to-noise ratio represent 18 total scenarios, for each of which we performed 10 simulations and model fittings. When fitting each model we utilized a Gamma(3,3) prior for the edge weights and Inverse-gamma(2,2) priors for $\sigma^2$ and $\tau$. Using the posterior samples we computed the average 90\% coverage rate for the posterior distribution of all edge weight parameters, along with RMSE and bias. The results of our simulation study can be found in Table \ref{simstudy}.

\begin{table}[ht]
\scriptsize
\centering
\begin{tabular}{rrrrrrrr}
  \hline
 Setting & Coverage & Bias & RMSE &Setting& Coverage & Bias & RMSE \\ 
  \hline
 $p=5^2, n= 10, \sigma^2= 0.2 $& 0.807 & -0.314 & 0.568 &$p=5^2, n= 10, \sigma^2= 0.8 $& 0.835 & -0.334 & 0.625 \\ 
  $p=5^2, n= 25, \sigma^2= 0.2 $ & 0.835 & -0.209 & 0.483 &$p=5^2, n= 25, \sigma^2= 0.8 $ & 0.868 & -0.297 & 0.603 \\ 
  $p=5^2, n= 50, \sigma^2= 0.2 $ & 0.818 & -0.177 & 0.435 &$p=5^2, n= 50, \sigma^2= 0.8 $ & 0.860 & -0.257 & 0.584 \\ 
  $p=10^2, n= 10, \sigma^2= 0.2 $ & 0.824 & -0.267 & 0.551 &$p=10^2, n= 10, \sigma^2= 0.8 $ & 0.857 & -0.314 & 0.610 \\ 
  $p=10^2, n= 25, \sigma^2= 0.2 $ & 0.834 & -0.222 & 0.495 &$p=10^2, n= 25, \sigma^2= 0.8 $ & 0.846 & -0.300 & 0.614 \\ 
  $p=10^2, n= 50, \sigma^2= 0.2 $ & 0.816 & -0.190 & 0.452 &$p=10^2, n= 50, \sigma^2= 0.8 $ & 0.857 & -0.277 & 0.593 \\ 
  $p=15^2, n= 10, \sigma^2= 0.2 $ & 0.819 & -0.284 & 0.563 &$p=15^2, n= 10, \sigma^2= 0.8 $ & 0.849 & -0.303 & 0.610 \\ 
  $p=15^2, n= 25, \sigma^2= 0.2 $ & 0.832 & -0.224 & 0.504 &$p=15^2, n= 25, \sigma^2= 0.8 $ & 0.843 & -0.310 & 0.616 \\ 
  $p=15^2, n= 50, \sigma^2= 0.2 $ & 0.841 & -0.176 & 0.435 &$p=15^2, n= 50, \sigma^2= 0.8 $ & 0.850 & -0.276 & 0.602 \\ 
   \hline
\end{tabular}
\caption{Average coverage rates, RMSE, and bias for all 18 settings in simulation study.}\label{simstudy}
\end{table}

From our simulation, we generally see improvement in average coverage, bias and RMSE as the number of samples increases, which is to be expected. Interestingly, the patterns relating to changes in lattice size or signal to noise ratio are less clear from our results. The consistent negative bias across simulation settings is potentially concerning, but is likely a result of the non-linear relationship between edge weights and covariance via the Mat\'{e}rn function. The fact that this bias decreases with sample size, along with our demonstration of model identifiability gives us confidence that this model functions in practice.

\section{Numerical Results}

\subsection{An Application Using NC eBird Data}

 Returning to the data discussed in the introduction and which motivated the development of this method, we now fit our model to bird abundance data within the state of North Carolina.  As alluded to in our introduction, there are varying patterns of between-site dependence among species within this spatial domain, and we are interested in obtaining the species-specific intrinsic distances which characterize that spatial dependence. The data were retrieved from the eBird database, a project of the Cornell Lab of Ornithology that allows bird watchers across the globe to record and submit their bird watching observations in the form of checklists that contain the specimen count for each observed species as well as the observer's location and ``effort," or time spent bird watching \citep{ebird}. The model was fit to the data for several common species within the state of North Carolina during the time period from January 2018 to December 2020. The species considered were the northern cardinal, the Carolina wren, the mourning dove, the turkey vulture, the Canada goose, the laughing gull, and the mallard. In addition to being abundant throughout the state, these species were chosen as they represent a considerable range of genetic and habitat diversity. 

 We treat each of the 100 counties of North Carolina as the nodes of a graph, with 256 edges connecting each pair of adjacent counties. For each species in the data set, observed counts were aggregated temporally by month and spatially by county. The data were stored in a $36 \times 100$ matrix $\bY$, in which $y_{ij}$ represents the combined number of individual birds from that species observed during month $i = 1,...,36$ within the borders of county $j = 1,...,100$ by all birdwatchers contributing to the database. In addition to the species-specific counts matrices, there is a $36 \times 100$ effort matrix $\bT$ common to all species, where $t_{ij}$ contains the total time spent bird watching during month $i$ within county $j$, as well as a geographic distance matrix $\bD^{Geo}$, obtained by calculating the physical distances between the geographical centroid of all bird watching locations within each county. We also have the $100 \times 3$ matrix of normalized county-level environmental covariates $\bX$ which contains the average county elevation, the percentage of county lands designated as urban vs. rural by the 2020 US Census, and percentage of county lands covered by water. Each of these covariates can influence the baseline abundance of various bird species throughout the region due to species' individual habitat preferences \citep{swick2016}.
 
 
 

 For each species, we model $\bY$ using the following generalized linear model with spatial random effects:

 \begin{equation}\label{cs_mod}
    \begin{aligned}
    y_{ij} &\sim \text{Poisson}(\text{exp}(\theta_{ij})t_{ij}) \\
    \theta_{ij} &= \beta_{0i} + x_j^\top \beta + u_{ij} + e_{ij} \\
    \bU &\sim N_{n\times p}(\bzero_{n \times p}, \bSigma(\sigma^2,\bW) \otimes \bI_n) \\
   \bE &\sim  N_{n\times p}(\bzero_{n \times p}, \psi \bI_{np}) \\
   \beta_0 &\in \mathbb{R}^n, \; \; \beta \in \mathbb{R}^3, \; \; \sigma^2,\psi >0, \; \; \bW \in \mathcal{W}_G.
\end{aligned}
\end{equation}

Here $\theta_{ij}$ is the log of the expected rate at which birds are observed during month $i$ within county $j$. The parameter $\beta_{0i}$ is a temporally-varying intercept term and $\beta$ the coefficient vector relating $x_j$, the vector of environmental covariates at county $j$ to the mean of $\theta_{ij}$. The spatial random effects matrix $\bU$ has independent rows and between-county covariance given by $\bSigma$ as defined as in equation \ref{model}; $\bE$ is a matrix of iid Gaussian noise. Of primary interest to us within this article is $\bSigma$ and the underlying distance matrix $\bD$, which can be viewed as containing the species-specific intrinsic distances between counties in North Carolina, with other model parameters primarily serving to enable an appropriate fit for $\bY$. We note that this model is structurally quite similar to the one proposed in \citet{bym1991} with the primary distinction being that a CAR model is used in their work to characterize the covariance of $\bU$.

We use the priors given in equation \ref{prior} for $\sigma^2$ and $\bW$ and use the following priors for the remaining model parameters:

\begin{equation}\label{more_priors}
\begin{aligned}
    \beta_{0} &\sim N(\bzero_n, \sigma^2_0\bI_n) \\
    \beta &\sim N(\bzero_3, \sigma^2_\beta\bI_3) \\
    \psi &\sim \text{Inverse-gamma}(a_{\psi},b_{\psi}) \\
    \sigma^2_{0} &\sim \text{Inverse-gamma}(a_{0},b_{0}) \\
    \sigma^2_{\beta} &\sim \text{Inverse-gamma}(a_{\beta},b_{\beta}).
\end{aligned}
\end{equation}
In our implementation of this model we set hyperprior parameters $a_{\sigma^2} = a_\psi = a_0 = a_\beta = 2$ and $b_{\sigma^2} = b_w = b_\psi = b_0 = b_\beta = 1$, choices made to be relatively uninformative. We note that we originally considered using an autoregressive prior for the vector of temporally-varying intercepts, but found that in practice it had very little impact on parameter estimates, but a fairly large negative on computational efficiency because doing so forces us to update $\theta$ by inverting an $np \times np$ matrix rather than a $p \times p$ matrix $n$ times. Additionally, because our interest is primarily in the structure of $\bSigma$ and the corresponding distance matrix, the degree of temporal correlation in $\beta_0$ is not important here. If our goal was to use this model for the purposes of forecasting, then more nuance would need to be applied to handling the spatiotemporal structure and interactions.

\subsection{Interpretation of Results}

The model is complex and the sampler used is somewhat computationally expensive, but we have found that it mixes well and within reasonable time periods for most small and medium sized graphs ($p < 500$ nodes). The data for the application described in the following section had $n = 36$ observations and $p = 100$ nodes. It required approximately eight hours to produce 10000 iterations of the sampler using code written in R and run single threaded on the lead author's personal laptop equipped with an Intel Core i7-9750H processor. The first 2500 iterations were treated as burn in and the 7500 we retained had an average effective sample size of approximately 3000, indicating reasonably good mixing. The code used for our analysis is included with the supplemental materials.

\begin{figure}[t!]
    \centering
    \includegraphics[width=10cm]{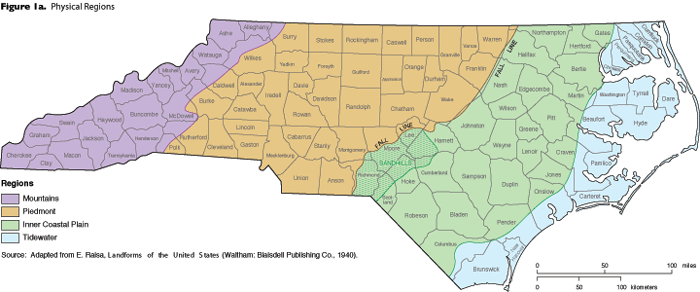}
    \caption{The four main geographic regions of the state of North Carolina. Image retrieved from NCpedia (ncpedia.org) and accessed on 20 February 2022.}
    \label{NC_geo}
\end{figure}

To interpret the output of our model, 
it is useful to have a cursory understanding of the geography of North Carolina. The majority of physical geographic variability within the state of North Carolina tends to run on an east to west axis. Figure \ref{NC_geo} depicts the four main geographic regions of the state, with the easternmost tidewater region, followed by the inner coastal plain, the Piedmont plateau in the center of the state, with the Blue Ridge mountains running through the westernmost part of North Carolina \citep{nc_geo,swick2016}. 

\begin{figure}[t!]
\centering
\subfloat[]{\includegraphics[width=.8\linewidth]{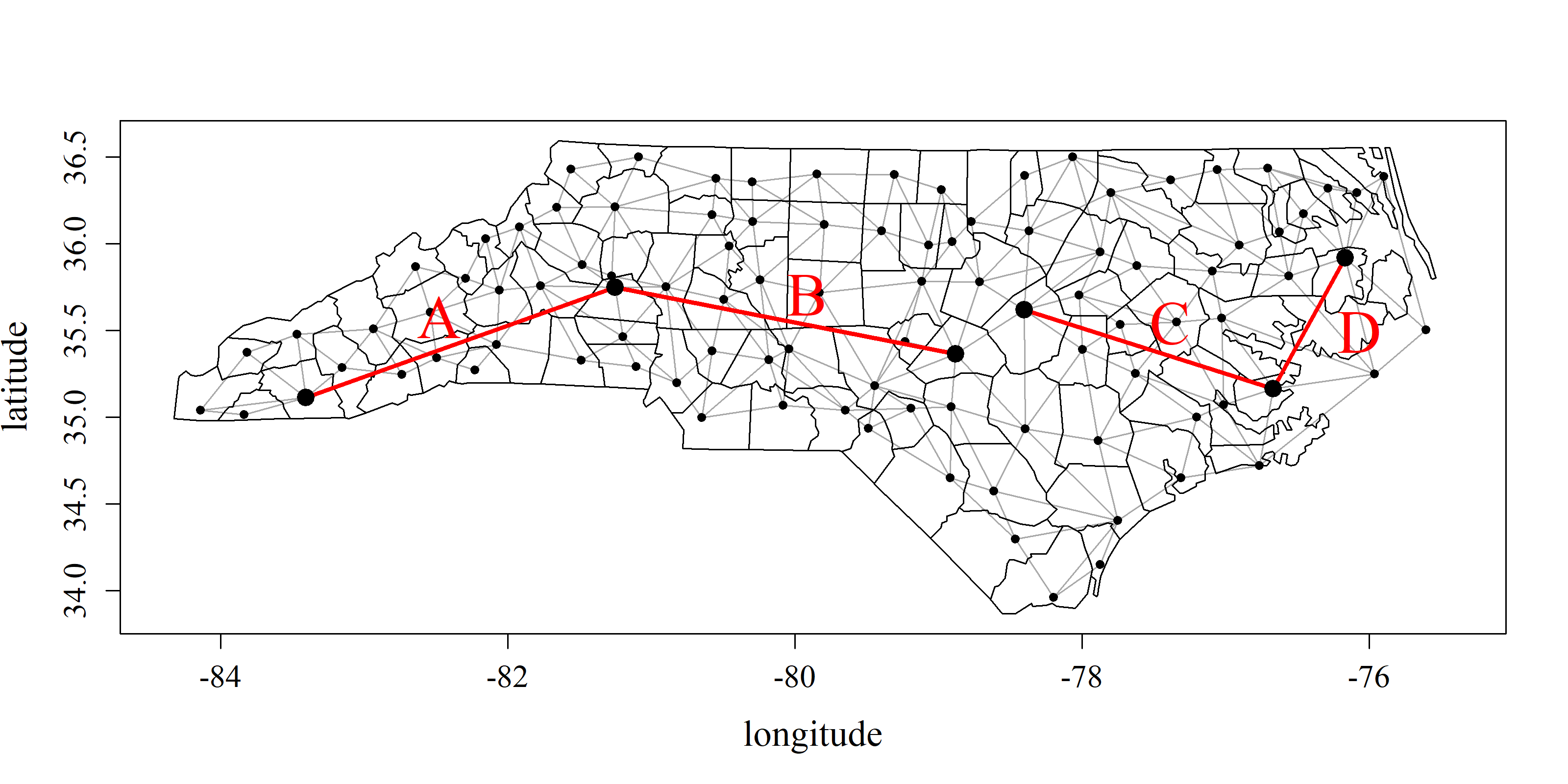}\label{post_a}}\par
\subfloat[]{\includegraphics[width=.5\linewidth]{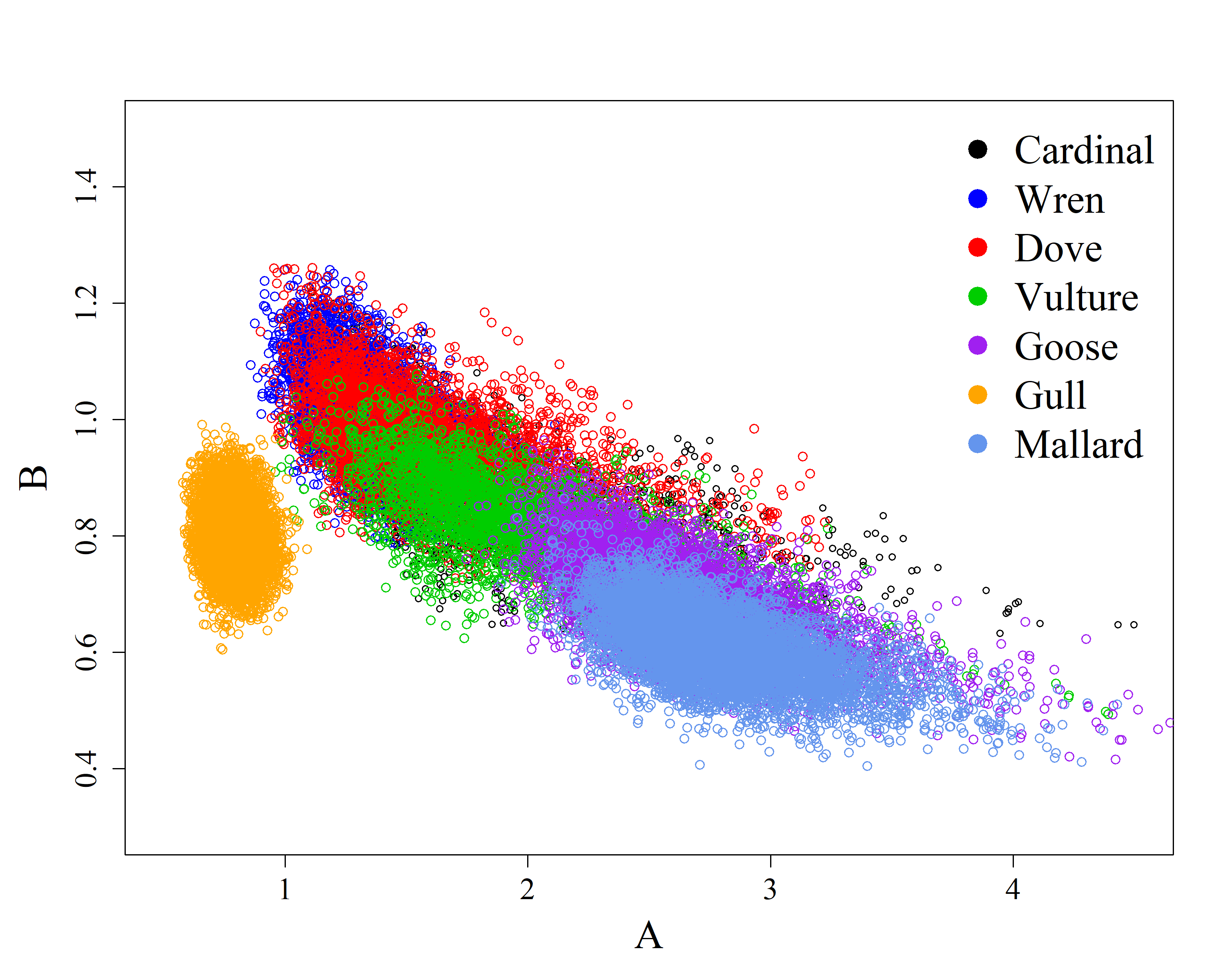}\label{post_b}}\hfill
\subfloat[]{\includegraphics[width=.5\linewidth]{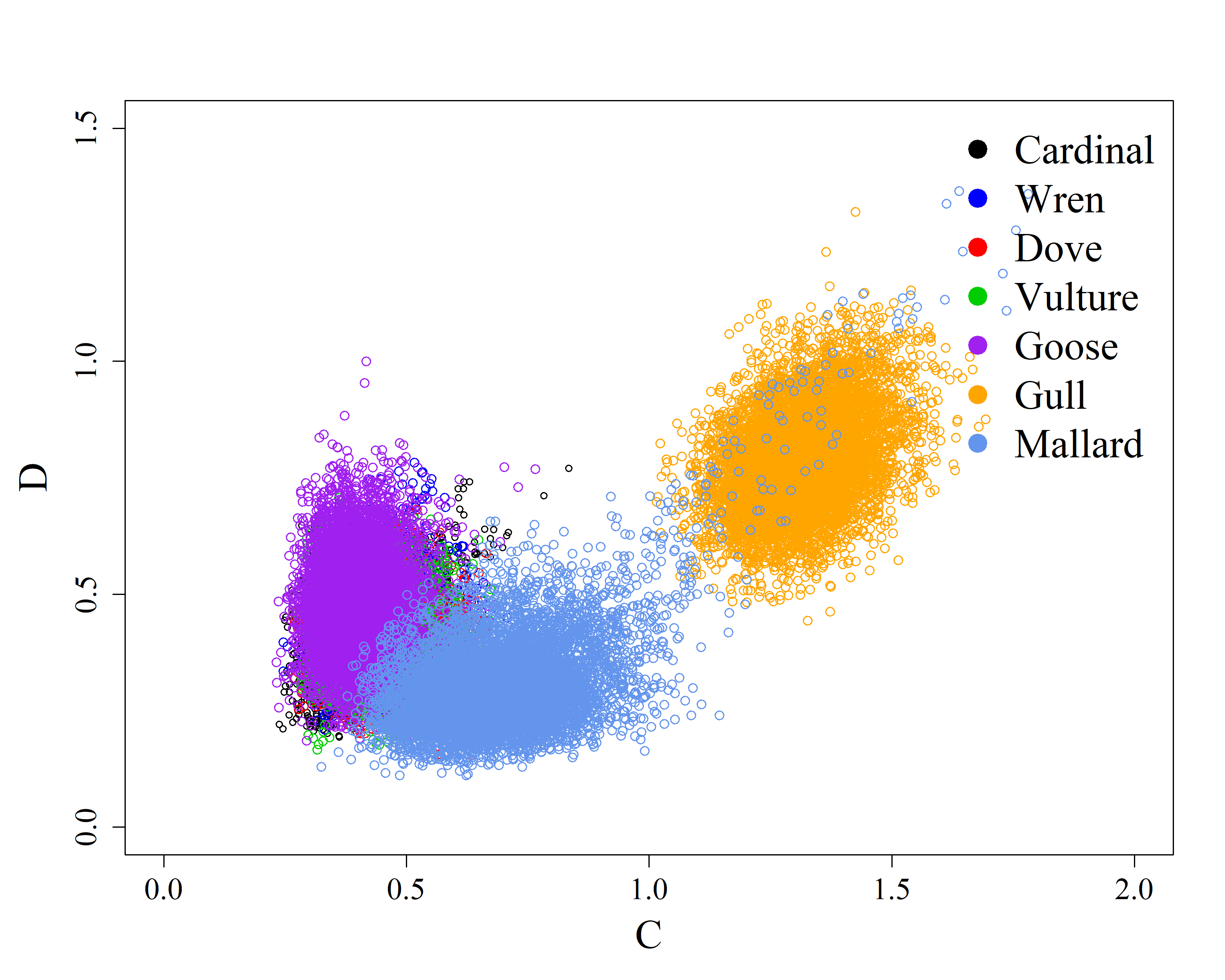}\label{post_c}}
\caption{Visualization of the uncertainty in posterior distances associated with four line segments for each bird species}
\label{post_comp}
\end{figure}

We are interested in how these geographic regions may manifest themselves within the intrinsic distances across the state for each species. Are there species for which distances along the eastern coast of the state are greater or smaller relative to the intrinsic distances from the coast moving inland? Are there species for which the mountain ranges in the west of the state significantly impact their spatial distribution while other species are unaffected? 
Figure \ref{post_comp} depicts the approximate posterior distribution for selected elements of the scaled posterior distance matrix for each species. Subfigure \ref{post_a} depicts the state of North Carolina and four line segments labeled ``A", ``B", ``C", and ``D", with these segments chosen to reflect various physical regions within the state, or the borders between them. For each of the seven bird species, there is a different distance associated with each of the four line segments; these distances are shorter, and the counties at their endpoints are ``closer" to one another if there is high correlation in the bird counts between locations. Subfigures \ref{post_b} and \ref{post_c} show the posterior distributions of the scaled intrinsic distances associated with two pairings of those segments. Both of these pairings corresponds to a geographic contrast that may be useful for understanding how these species spatial distributions interact with the physical properties of the environments they inhabit.  We see that the intrinsic distances associated with each of the segments differ from species to species, with the posterior for the laughing gull being more divergent from the other six species in both subfigures. This divergence makes sense when one considers that the laughing gull it is the only seabird of the seven species, having a natural range of habitats with the least overlap among the species considered \citep{swick2016}. We see this in subfigure \ref{post_c}, which highlights the contrast between intrinsic distances running along the coast (segment ``D") versus ones moving inland (segment ``C"). Subfigure \ref{post_b} highlights the contrast between the distances associated with segment ``A," which spans across a mountain range, and ``B" which runs along the relatively flat Piedmont. The spatial distributions of birds with posteriors concentrated at higher values along the x-axis of this plot, such as the Canada goose and Mallard are more impacted by the changes in elevation and habitat along segment ``A" than the others.

We would also like to visualize the entire mean posterior distance matrix for each species. Because this matrix corresponds to a high dimensional Euclidean embedding of the graph, it is challenging to represent with a single plot and without significant distortion. To this end, we present a visualization which highlights the contrast between the posterior mean intrinsic distance and the physical distance between the centroids of each pair of adjacent counties, which in the minds of the authors is more interpretable than the raw edge weights themselves; this  visualization provides us not only a sense of the relative distances between counties, but more significantly of the deformation of the graph's geographic orientation necessary to produce the intrinsic distances. We have done this by calculating $z^d_{jk}$, a score designed to represent whether the intrinsic distance between adjacent nodes $j$ and $k$ is smaller or greater than suggested by the physical distance between those nodes.   
\begin{equation}
     z^d_{jk} = Z(\bD^{Geo}(E)) -  Z(\overline{(\bD(E)|\bY)})_{jk}
\end{equation}
Here $\bD(E)$ indicates the set of distances between nodes connected by edges of our graph, and $\overline{(\bD(E)|\bY)})$ is the set of posterior means for those same distances. The function $Z(\cdot)$ scales a set to have mean zero and standard deviation one. High values of $z^d_{jk}$ indicate that nodes $j$ and $k$ are intrinsically ``closer" together relative to their physical separation, while low values indicate that they are more distant. 

\begin{figure}[t!]
    \centering
    \subfloat[]{\includegraphics[width=12cm]{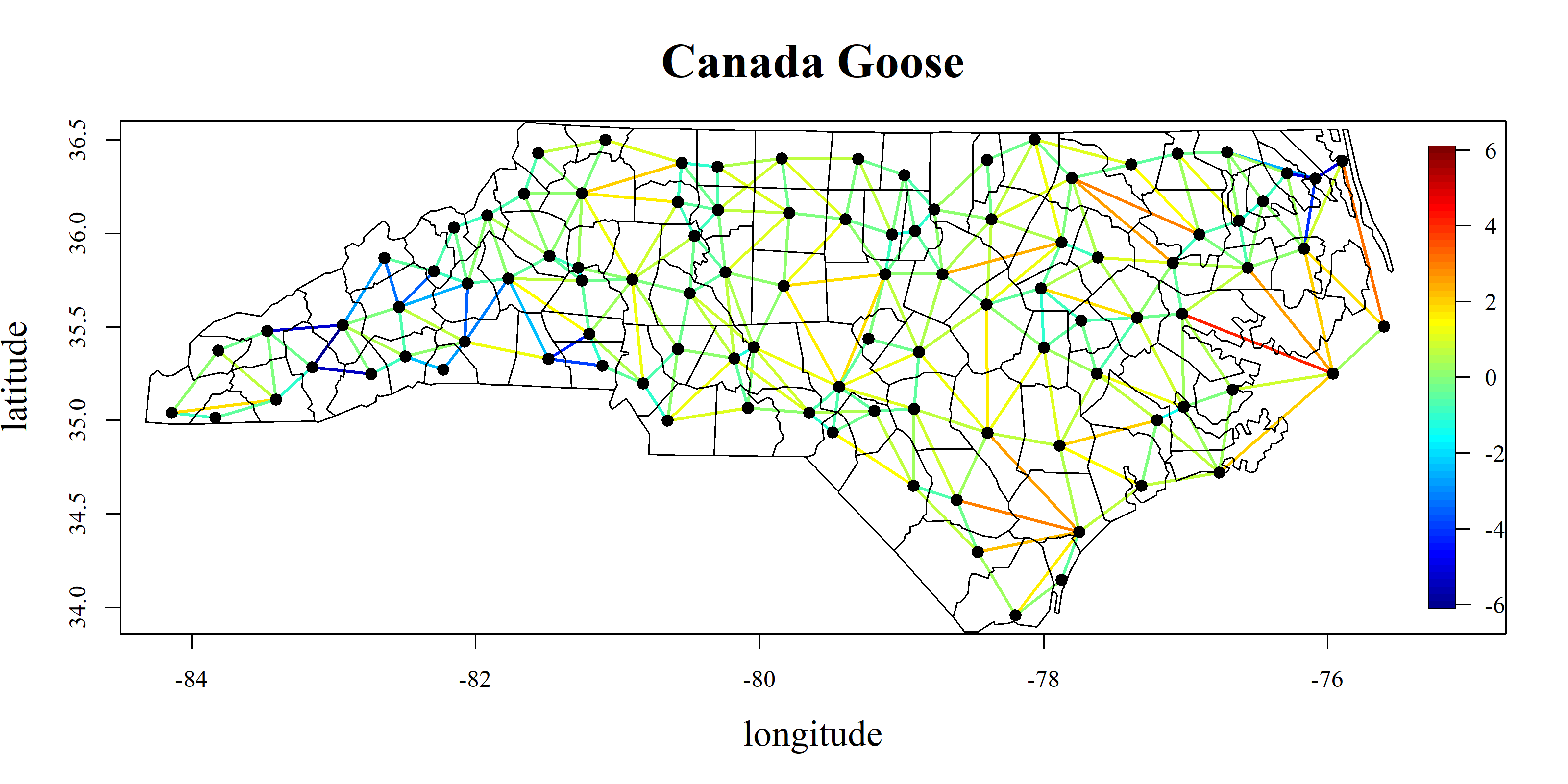}\label{goose}} \par
    \subfloat[]{\includegraphics[width=12cm]{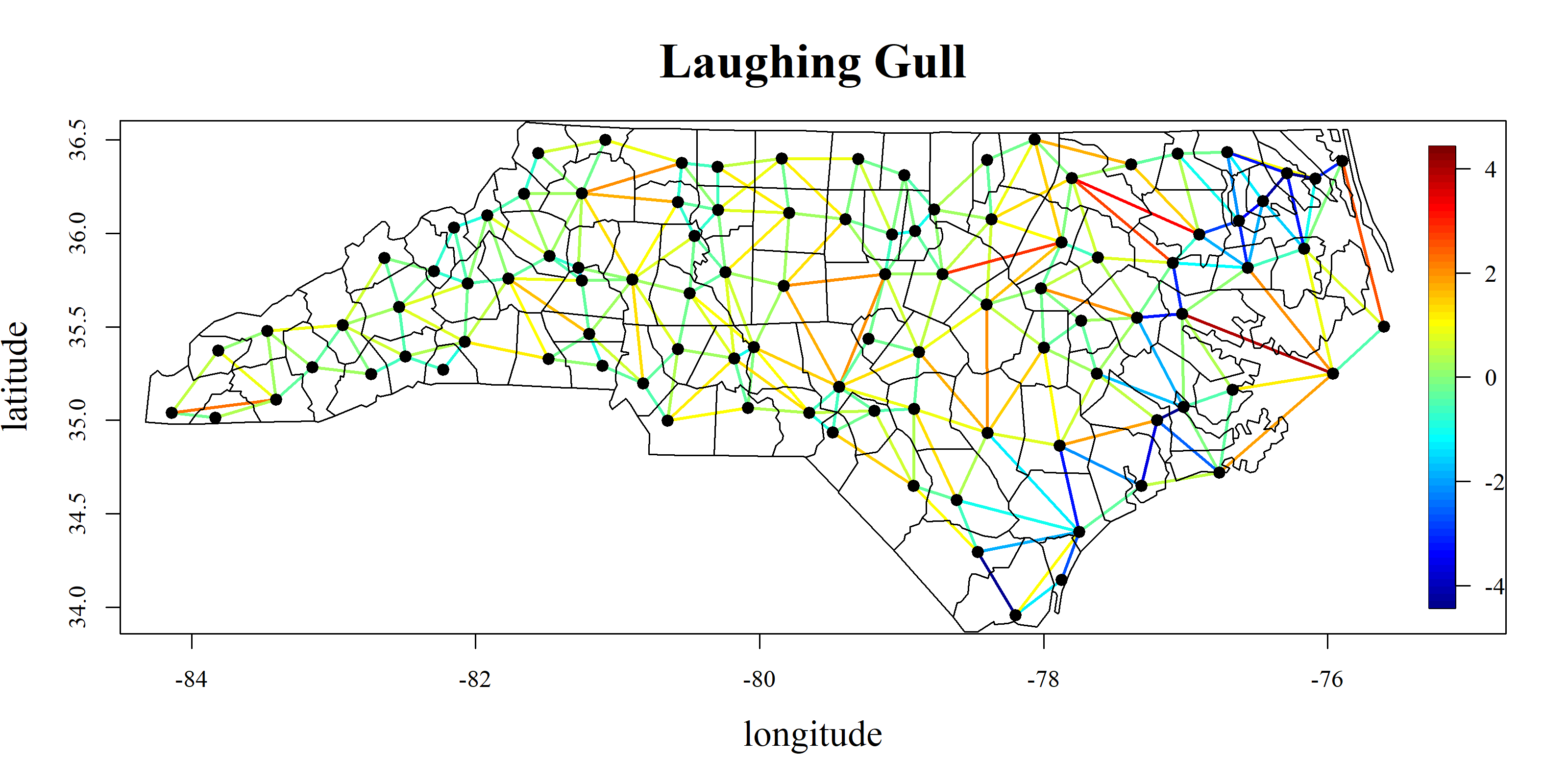}\label{lgull}}
    \caption{Maps of $\bz^d$ for Mallard and Laughing Gull. Warm colors indicate areas which are ``closer" than suggested by their geographic distance and cool colors more distant.}
    \label{full_vis}
\end{figure}

Figure \ref{full_vis} contains two plots depicting $z^d_{jk}$ for all pairs of adjacent counties. Subfigure \ref{goose} contains a map of North Carolina for the Canada goose, a migratory species which can be observed year round within the state, but is considerably more common during the winter \citep{swick2016}. We note the bands of  dark blue line segments running approximately along longitude $-82^\circ E$, separating the westernmost portion of the state from its central regions, which indicates that observations of Canada geese on opposite sides of this border are less correlated than ones made on the same side. Significantly, that border corresponds well to the mountain ranges making up the westernmost parts of North Carolina. Ornithologists have established that mountain ranges present significant barriers to the flight paths of migratory birds \citep{aschwanden2020} which leads to a physical interpretation for the posterior correlations identified by our model.

Subfigure \ref{lgull} depicts a comparable map for the laughing gull, which is found most abundantly along the coast of the state, and becomes progressively less common as one moves inland \citep{swick2016}. While many of the edges between coastal counties are red and orange, indicating greater connectivity in the regions, the majority of the edges connecting the coastal counties bordering the Albermale sound (at $-76^\circ E, 36^\circ N$) are blue and green, indicating that not all coastal locations possess the same level of interdependence. We also note the existence of a blue band of edges running parallel to the coast at approximately $-77^\circ E$, along the border between the coast and coastal plains regions of the state. This could be viewed as a boundary between coastal and inland regions over which laughing gull populations tend not to cross, and which is revealed within the posterior distance matrix of our model. 

For both species, these plots reveal significant interactions between their spatial distributions and the geography of where they were observed, even after having accounted for how environmental covariates may impact their baseline abundance rates within each county. Taken together, the plots (along with the plots not included for the other five species analyzed) reveal that there is a unique covariance structure and underlying distance between locations associated with each species. 

An additional quantity of interest in many spatial data applications is the ratio $\sigma^2/(\sigma^2+\psi)$ which may be thought of as the percentage of total variability attributable to the spatial process, in comparison to the nugget effect represented by $\psi$. Table \ref{ratio} contains a posterior summary of this ratio for each of the seven bird species. We note that it is particularly low for the Carolina wren and mourning dove, indicating weak spatial correlation in the data for county-level populations of those particular species.

\begin{table}[t!]
\centering
\begin{tabular}{rll}
  \hline
 & Mean & 95\% CI \\ 
  \hline
Northern Cardinal & 0.584 & (0.429,0.712) \\ 
  Carolina Wren & 0.064 & (0.014,0.298) \\ 
  Mourning Dove & 0.129 & (0.058,0.313) \\ 
  Turkey Vulture & 0.371 & (0.262,0.476) \\ 
  Canada Goose & 0.422 & (0.312,0.693) \\ 
  Laughing Gull & 0.857 & (0.789,0.897) \\ 
  Mallard & 0.471 & (0.400,0.543) \\ 
   \hline
\end{tabular}
\caption{Posterior mean and 95\% credible intervals of ratio $\sigma^2/(\sigma^2+\psi)$ for all seven analyzed species}
\label{ratio}
\end{table}

\section{Discussion}

We now briefly discuss a few potential applications, extensions, and questions related to the ideas presented within this article. One of the most significant pieces of output from our model is the posterior distance matrix. By itself, it can be analyzed to better understand the relationships between nodes of a graph in terms of their latent proximity to one another; but as was the case in our application with the eBird data, there are instances in which one may fit this model separately for multiple datasets (unique bird species in the case of the eBird example) which are observed on the same graph. In such cases, an analysis of how the posterior distance matrices for each dataset compare to one another may reveal significant information about the relationships between observed variables. \citet{abdi2005} presents a framework for analyzing multiple distance matrices computed on the same set of areal units. Such a framework could be used for example to perform a clustering analysis of common bird species in North Carolina based on their unique distance matrices. 

When modeling the eBird data, we included environmental covariates in the mean of our spatial process. An alternative or complimentary approach could be to include them in the specification of the model's edge weights, which would enhance the interpretability of the model at the edge weights level, and not just in terms of the distance matrix. One could imagine based on the findings illustrated in Subfigure \ref{goose} that edge weights for the Canada goose may be negatively associated with elevation. An approach along these lines was explored using CAR models by \citet{hanks2013} and \citet{wang2016}, and could be augmented within our model framework by treating edge weights as a linear combination of spatial covariates and an additional spatial random effect over the line graph of our network. Additionally, species count observations are often zero-inflated. Accounting for this as in \citet{agarwal2002} could prove beneficial to model quality in applied settings. Other potential extensions include joint species modeling using our methodology. One approach could be to perform factor analysis \citep{lawley1962, harman1976} on the model parameters characterizing species-specific covariance 
enabling one to better understand which species interact with the spatial domain similarly in terms of covariance and intrinsic distances, while providing a natural approach to modelling the intrinsic distances for each species hierarchically.

The method described in this article provides a flexible and novel framework for modeling the covariance of data observed on a graph structure. It's use of an estimated  distance metric to which standard covariance functions may be applied allows for interpretation more similar to that of models in continuous space settings as compared to CAR models. We have demonstrated that our model is flexible enough to represent covariance structures in which patterns of spatial autocorrelation vary throughout the domain, and that there are contexts in which it performs better than advanced versions of autoregressive models which are the traditional tools for modeling spatially dependent graph data. We also find the specification of between-node covariance in terms of intrinsic distances a useful framing for understanding and interpreting model results, as exemplified by our discussion of model output in section 4.2 in which we found that an analysis of the posterior distance matrix revealed connections between the covariance in the eBird dataset and the physical geography of the spatial domain on which it was observed.

\section*{Supplemental Materials}

Additional information and supporting material for this article is available online at the journal's website.

\bibliographystyle{unsrtnat}
\bibliography{ref}

\begin{thebibliography}{47}
\providecommand{\natexlab}[1]{#1}
\providecommand{\url}[1]{\texttt{#1}}
\expandafter\ifx\csname urlstyle\endcsname\relax
  \providecommand{\doi}[1]{doi: #1}\else
  \providecommand{\doi}{doi: \begingroup \urlstyle{rm}\Url}\fi

\bibitem[Ver~Hoef et~al.(2018{\natexlab{a}})Ver~Hoef, Peterson, Hooten, Hanks,
  and Fortin]{verhoef2018a}
Jay~M. Ver~Hoef, Erin~E. Peterson, Mevin~B. Hooten, Ephraim~M. Hanks, and
  Marie-Jos\'{e}e Fortin.
\newblock Spatial autoregressive models for statistical inference from
  ecological data.
\newblock \emph{Ecological Monographs}, 88\penalty0 (1):\penalty0 36--59,
  2018{\natexlab{a}}.

\bibitem[Guttorp et~al.(1994)Guttorp, Meiring, and Sampson]{guttorp1994}
Peter Guttorp, Wendy Meiring, and Paul~D. Sampson.
\newblock A space-time analysis of ground-level ozone data.
\newblock \emph{Environmetrics}, 5\penalty0 (3):\penalty0 241--254, 1994.

\bibitem[Sullivan et~al.(2009)Sullivan, Wood, Iliff, Bonney, Fink, and
  Kelling]{ebird}
Brian~L. Sullivan, Christopher~L. Wood, Marshall~J. Iliff, Rick~E. Bonney,
  Daniel Fink, and Steve Kelling.
\newblock ebird: A citizen-based bird observation network in the biological
  sciences.
\newblock \emph{Biological conservation}, 142\penalty0 (10):\penalty0
  2282--2292, 2009.

\bibitem[{Coastal Area Management Act}(1974)]{cama1974}
{Coastal Area Management Act}.
\newblock N.C.G.S \S 113A-103 (2), 1974.

\bibitem[Fisher(1921)]{fisher1921}
Ronald~A. Fisher.
\newblock 014: On the ``probable error" of a coefficient of correlation deduced
  from a small sample.
\newblock 1921.

\bibitem[Besag(1974)]{besag1974}
Julian Besag.
\newblock Spatial interaction and the statistical analysis of lattice systems.
\newblock \emph{Journal of the Royal Statistical Society: Series B
  (Methodological)}, 36\penalty0 (2):\penalty0 192--225, 1974.

\bibitem[Besag et~al.(1991)Besag, York, and Molli\'{e}]{bym1991}
Julian Besag, Jeremy York, and Annie Molli\'{e}.
\newblock Bayesian image restoration with two applications in spatial
  statistics.
\newblock \emph{Annals of the institute of statistical mathematics},
  43\penalty0 (1):\penalty0 1--20, 1991.

\bibitem[Gelfand and Vounatsou(2003)]{gelfand2003}
Alan~E. Gelfand and Penelope Vounatsou.
\newblock Proper multivariate conditional autoregressive models for spatial
  data analysis.
\newblock \emph{Biostatistics}, 41\penalty0 (1):\penalty0 11--15, 2003.

\bibitem[Hanks and Hooten(2013)]{hanks2013}
Ephraim~M. Hanks and Mevin~B. Hooten.
\newblock Circuit theory and model-based inference for landscape connectivity.
\newblock \emph{Journal of the American Statistical Association}, 108\penalty0
  (501):\penalty0 22--33, 2013.

\bibitem[Ejigu and Wencheko(2020)]{ejigu2020}
Bedilu~A. Ejigu and Eshetu Wencheko.
\newblock Introducing covariate dependent weighting matrices in fitting
  autoregressive models and measuring spatio-environmental autocorrelation.
\newblock \emph{Spatial Statistics}, 38:\penalty0 100454, 2020.

\bibitem[Ver~Hoef et~al.(2018{\natexlab{b}})Ver~Hoef, Hanks, and
  Hooten]{verhoef2018b}
Jay~M. Ver~Hoef, Ephraim~M. Hanks, and Mevin~B. Hooten.
\newblock On the relationship between conditional {(CAR)} and simultaneous
  {(SAR)} autoregressive models.
\newblock \emph{Spatial statistics}, 25:\penalty0 68--85, 2018{\natexlab{b}}.

\bibitem[Gramacy and Apley(2015)]{gramacy2015}
Robert~B. Gramacy and Daniel~W. Apley.
\newblock Local gaussian process approximation for large computer experiments.
\newblock \emph{Journal of Computational and Graphical Statistics}, 24\penalty0
  (2):\penalty0 561--578, 2015.

\bibitem[Katzfuss and Guinness(2021)]{katzfuss2021}
Matthias Katzfuss and Joseph Guinness.
\newblock A general framework for becchia approximations of gaussian processes.
\newblock \emph{Statistical Science}, 36\penalty0 (1):\penalty0 124--141, 2021.

\bibitem[Guinness(2018)]{guinness2018}
Joseph Guinness.
\newblock Permutation and grouping methods for sharpening gaussian process
  approximations.
\newblock \emph{Technometrics}, 60\penalty0 (4):\penalty0 415--429, 2018.

\bibitem[Lee(2011)]{lee2011}
Duncan Lee.
\newblock A comparison of conditional autoregressive models used in bayesian
  disease mapping.
\newblock \emph{Spatial and Spatio-temporal Epidemiology}, 2\penalty0
  (2):\penalty0 79--89, 2011.

\bibitem[Hughes and Haran(2013)]{hughes2013}
John Hughes and Murali Haran.
\newblock Dimension reduction and alleviation of confounding for spatial
  generalized linear mixed models.
\newblock \emph{Journal of the Royal Statistical Society: Series B (Statistical
  Methodology)}, 75\penalty0 (1):\penalty0 139--159, 2013.

\bibitem[Cressie(1993)]{cressie1993}
Noel Cressie.
\newblock \emph{Statistics for spatial data}.
\newblock John Wiley \& Songs, 1993.

\bibitem[Sampson and Guttorp(1992)]{sampson1992}
Paul~D. Sampson and Peter Guttorp.
\newblock Nonparametric estimation of nonstationary spatial covariance
  structure.
\newblock \emph{Journal of the American Statistical Association}, 87\penalty0
  (417):\penalty0 108--119, 1992.

\bibitem[Schmidt and O'Hagan(2003)]{schmidt2003}
Alexandra~M. Schmidt and Anthony O'Hagan.
\newblock Bayesian inference for nonstationary spatial covariance structures
  via spatial deformations.
\newblock \emph{Journal of the Royal Statistical Society: Series B (Statistical
  Methodology)}, 65\penalty0 (3):\penalty0 743--758, 2003.

\bibitem[Bornn et~al.(2012)Bornn, Shaddick, and Zidek]{bornn2012}
Luke Bornn, Gavin Shaddick, and James~V. Zidek.
\newblock Modeling nonstationary processes through dimension expansion.
\newblock \emph{Journal of the American Statistician}, 107\penalty0
  (497):\penalty0 281--289, 2012.

\bibitem[Mat\'{e}rn(1960)]{matern1960}
Bertil Mat\'{e}rn.
\newblock Spatial variation - stochastic models and their applications to some
  problems in forest survey sampling investigations.
\newblock \emph{Report of the Forest Research Institute of Sweden}, 1960.

\bibitem[Banerjee et~al.(2003)Banerjee, Carlin, and Gelfand]{banerjee2003}
Sudipto Banerjee, Bradley~P. Carlin, and Alan~E. Gelfand.
\newblock \emph{Hierarchical modeling and analysis for spatial data}.
\newblock Chapman and Hall/CRC, 2003.

\bibitem[Ver~Hoef(2018)]{verhoef2018c}
Jay~M. Ver~Hoef.
\newblock Kriging models for linear networks and non-euclidean distances:
  Cautions and solutions.
\newblock \emph{Methods in Ecology and Evolution}, 9\penalty0 (6):\penalty0
  1600--1613, 2018.

\bibitem[Gihman and Skorohod(1974)]{gihman1974}
Iosif~I. Gihman and Anatolii~V. Skorohod.
\newblock \emph{The theory of stochastic processes}.
\newblock Berlin: Springer-Verlag, 1974.

\bibitem[Jungnickel(2012)]{jungnickel2012}
Dieter Jungnickel.
\newblock \emph{Graphs, networks and algorithms}.
\newblock Berlin: Springer, 2012.

\bibitem[Chebotarev(2011)]{chebotarev2011}
Pavel Chebotarev.
\newblock A class of graph-geodetic distances generalizing the shortest-path
  and the resistance distances.
\newblock \emph{Discrete Applied Mathematics}, 159\penalty0 (5):\penalty0
  295--302, 2011.

\bibitem[Klein and Zhu(1998)]{klein1998}
Douglas~J. Klein and H-Y. Zhu.
\newblock Distances and volumina for graphs.
\newblock \emph{Journal of mathematical chemistry}, 23\penalty0 (1):\penalty0
  179--195, 1998.

\bibitem[Klein and Randi\'{c}(1993)]{klein1993}
Douglas~J. Klein and Milan Randi\'{c}.
\newblock Resistance distance.
\newblock \emph{Journal of mathematical chemistry}, 12\penalty0 (1):\penalty0
  81--95, 1993.

\bibitem[Chandra et~al.(1996)Chandra, Raghavan, Ruzzo, Smolensky, and
  Tiwari]{chandra1996}
Ashok~K. Chandra, Prabhakar Raghavan, Walter~L. Ruzzo, Roman Smolensky, and
  Prasoon Tiwari.
\newblock The electrical resistance of a graph captures its commute and cover
  times.
\newblock \emph{Computational complexity}, 6\penalty0 (4):\penalty0 312--340,
  1996.

\bibitem[Thiele et~al.(2018)Thiele, Buchholz, and Schirmel]{thiele2018}
Jan Thiele, Sascha Buchholz, and Jens Schirmel.
\newblock Using resistance distance from circuit theory to model dispersal
  through habitat corridors.
\newblock \emph{Journal of Plant Ecology}, 11\penalty0 (3):\penalty0 385--393,
  2018.

\bibitem[Peterson et~al.(2019)Peterson, Hanks, Ver~Hoef, Hooten, and
  Fortin]{peterson2019}
Erin~E. Peterson, Ephraim~M. Hanks, Jay~M. Ver~Hoef, Mevin~B. Hooten, and
  Marie-Jos\'{e}e Fortin.
\newblock Intrinsic graph distances compared to euclidean distances for
  correspondent graph embedding.
\newblock \emph{Ecological Monographs}, 89\penalty0 (2):\penalty0 e01355, 2019.

\bibitem[Zhu and Klein(1996)]{zhu1996}
H-Y. Zhu and Douglas~J. Klein.
\newblock Graph-geometric invariants for molecular structures.
\newblock \emph{Journal of chemical information and computer sciences},
  36\penalty0 (6):\penalty0 1067--1075, 1996.

\bibitem[Ivanciuc et~al.(2001)Ivanciuc, Ivanciuc, and Klein]{ivanciuc2001}
Ovidiu Ivanciuc, Teodora Ivanciuc, and Douglas~J. Klein.
\newblock Intrinsic graph distances compared to euclidean distances for
  correspondent graph embedding.
\newblock \emph{MATCH Communications in Mathematical and in Computer
  Chemistry}, 44:\penalty0 251--278, 2001.

\bibitem[White and Ghosh(2009)]{white2009}
Gentry White and Sujit~K. Ghosh.
\newblock A stochastic neighborhood conditional autoregressive model for
  spatial data.
\newblock \emph{Computational Statistics and Data Analysis}, 53\penalty0
  (8):\penalty0 3033--3046, 2009.

\bibitem[Ma et~al.(2010)Ma, Carlin, and Banerjee]{ma2010}
Haijun Ma, Bradley~P. Carlin, and Sudipto Banerjee.
\newblock Hierarchical and joint site-edge methods for medicare hospice service
  region boundary analysis.
\newblock \emph{Biometrics}, 66\penalty0 (2):\penalty0 355--364, 2010.

\bibitem[Stein(1999)]{stein1999}
Michael~L. Stein.
\newblock \emph{Interpolation of spatial data: some theory for kriging}.
\newblock Springer Science \& Business Media, 1999.

\bibitem[Smith et~al.(2015)Smith, Wakefield, and Dobra]{smith2015}
Thersa~R. Smith, Jon Wakefield, and Adrian Dobra.
\newblock Restricted covariance priors with applications in spatial statistics.
\newblock \emph{Bayesian analysis (Online)}, 10\penalty0 (4):\penalty0 965,
  2015.

\bibitem[Wang et~al.(2016)Wang, Yang, Lee, Ji, and You]{wang2016}
Xuesong Wang, Junguang Yang, Chris Lee, Zhuoran Ji, and Shikai You.
\newblock Macro-level safety analysis of pedestrian crashes in shanghai, china.
\newblock \emph{Accident Analysis \& Prevention}, 96:\penalty0 12--21, 2016.

\bibitem[Zellner(1988)]{zellner1988}
Arnold Zellner.
\newblock Optimal information processing and bayes's theorem. with comments and
  a reply by the author.
\newblock \emph{American Statistician}, 42\penalty0 (4):\penalty0 278--284,
  1988.

\bibitem[Gelman et~al.(2014)Gelman, Hwang, and Vehtari]{gelman2014}
Andrew Gelman, Jessica Hwang, and Aki Vehtari.
\newblock Understanding predictive information criteria for bayesian models.
\newblock \emph{Statistics and computing}, 24\penalty0 (6):\penalty0 997--1016,
  2014.

\bibitem[Swick(2016)]{swick2016}
Nate Swick.
\newblock \emph{American Birding Association Field Guide to Birds of the
  Carolinas}.
\newblock Scott \& Nix Inc., 2016.

\bibitem[NCpedia(2015)]{nc_geo}
NCpedia.
\newblock Our state geography in a snap: three regions overview --- {NC}pedia,
  2015.
\newblock URL \url{http://www.ncpedia.org/our-state-geography-snap-three}.
\newblock [Online; accessed 20-February-2022].

\bibitem[Aschwanden et~al.(2020)Aschwanden, Schmidt, Wichmann, and
  et~al.]{aschwanden2020}
Janine Aschwanden, Matthias Schmidt, Gabor Wichmann, and et~al.
\newblock Barrier effects of mountain ranges for broad-front bird migration.
\newblock \emph{Journal of Ornithology}, 161:\penalty0 59--71, 2020.

\bibitem[Abdi et~al.(2005)Abdi, O'Toole, Valentin, and Edelman]{abdi2005}
Herv\'{e} Abdi, Alice~J. O'Toole, Dominique Valentin, and Betty Edelman.
\newblock {DISTATIS: T}he analysis of multiple distance matrices.
\newblock \emph{2005 IEEE Computer Society Conference on Computer Vision and
  Pattern Recognition (CVPR'05)-Workshops}, pages 42--44, 2005.

\bibitem[Agarwal et~al.(2002)Agarwal, Gelfand, and Citron-Pousty]{agarwal2002}
Deepak~K. Agarwal, Alan~E. Gelfand, and Steven Citron-Pousty.
\newblock Zero-inflated models with application to spatial count data.
\newblock \emph{Environmental and Ecological Statistics}, 9\penalty0
  (4):\penalty0 341--355, 2002.

\bibitem[Lawley and Maxwell(1962)]{lawley1962}
David~N. Lawley and Adam~E. Maxwell.
\newblock Factor analysis as a statistical method.
\newblock \emph{Journal of the Royal Statistical Society. Series D (The
  Statistician)}, 12\penalty0 (3):\penalty0 209--229, 1962.

\bibitem[Harman(1976)]{harman1976}
Harry~H. Harman.
\newblock \emph{Modern factor analysis}.
\newblock University of Chicago press, 1976.

\end{thebibliography}

\end{document}


\setstretch{1.7}
\maketitle
\vspace{-2em}

\section{Proof of Identifiability}
\begin{numprop}{1}[Identifiability]
Using the parameterization for $\bSigma$ given in Equation \ref{modelap}, $(\sigma^2_1, \bW_1) \neq  (\sigma^2_2, \bW_2) \Rightarrow \bSigma_1 \neq \bSigma_2$ for all $\sigma^2_1,\sigma^2_2 > 0$ and all $\bW_1, \bW_2 \in \mathcal{W}_G$.
\end{numprop}

The equation relating $\bSigma$ to model parameters $(\sigma^2,\bW)$ is provided again for reference:

\begin{equation}\label{modelap}
\begin{aligned}
    \Sigma_{jk} &= \sigma^2 \rho_\nu(d_{ij}) \\
    d_{jk} &= \sqrt{(\be_j - \be_k)^\top\{\bL^+\}^2(\be_j-\be_k)} \\
    \bL &= \text{diag}(\bW \bone_p) - \bW 
\end{aligned}
\end{equation}

Because $\rho_\nu(\cdot)$ is a correlation function, and the distance from any node to itself is zero, the diagonal elements of $\bSigma$ are equal to $\sigma^2$. Thus $\sigma^2_1 \neq \sigma^2_2 \Rightarrow \bSigma_1 \neq \bSigma_2$. For two distance matrices $\bD_1 \neq \bD_2$ there exists some pair of nodes $(j,k)$ such that $d_{1jk} \neq d_{2jk}$. Because $\rho_\nu(\cdot)$ is a strictly decreasing function, $d_{1jk} \neq d_{2jk} \Rightarrow \Sigma_{1jk} \neq \Sigma_{2jk}$. 

We now need only to prove that $\bW_1 \neq \bW_2 \Rightarrow \bD_1 \neq \bD_2$. The transformation from edge weights matrix to to distance matrix as defined by the quasi-Euclidean metric can be rewritten as follows:

\begin{equation}
\begin{aligned}
\bD  &= \left( \bone_p \bd_{\{\bL^+\}^2}^\top  + \bd_{\{\bL^+\}^2} \bone_p^\top - 2\{\bL^+\}^2 \right)^{\circ \frac{1}{2}} \\
\bL &= \text{diag}(\bW \bone_p) - \bW
\end{aligned}
\end{equation}
Here $\bd_{\{\bL^+\}^2}$ is defined to be the vector with elements equal to the diagonal entries of $\{\bL^+\}^2$ and $(\cdot)^{\circ \frac{1}{2}}$ denotes the Hadamard (element-wise) square root of a matrix. The Hadamard root is an invertible transformation ($\bB_1 \neq \bB_2 \Leftrightarrow \bB_1^{\circ \frac{1}{2}} \neq \bB_2^{\circ \frac{1}{2}}$ for all matrices $\bB_1,\bB_2$) but the transformation $\bB = \Delta(\bA)$ for any $p \times p$ matrix $\bA$, given below, is not invertible.

\begin{equation}\label{transform}
    \bB = \Delta(\bA) = \bone\bd^\top_\bA + \bd_\bA \bone^\top - 2\bA
\end{equation}

This transformation is fairly well known in other statistical applications involving distances, and appears within the formulation of the quasi-Euclidean metric. A linear transformation $f(\cdot)$ is injective iff its null space $\mathcal{N}(f) =\{\bzero\}$. The null space of $\Delta$ is $\mathcal{N}(\Delta) = \{\bA: \bA = \bone \bc^\top + \bc \bone^\top, \bc \in \mathbb{R}^p\}$ as demonstrated below:

\begin{equation}
    \begin{aligned}
         &\text{Suppose } \Delta(\bA) = \bzero_{p \times p}: \\
         &\Rightarrow \bA = \bone\bd'_\bA/2 + \bd_\bA \bone'/2 \\
         &\forall \bc \in \mathbb{R}^p, \text{ let } \bA = \bone\bc^\top + \bc\bone^\top \\ 
         &\Rightarrow \bd_\bA = 2\bc \\
         &\Rightarrow \bA = \bone\bd'_\bA/2 + \bd_\bA \bone'/2 \\
         & \Rightarrow \mathcal{N}(\Delta) = \{\bA: \bA = \bone \bc^\top + \bc \bone^\top, \bc \in \mathbb{R}^p\}
    \end{aligned}
\end{equation}

Let $\mathcal{A}_p$ be the space of $p \times p$ symmetric matrices such that $\forall \bA \in \mathcal{A}_p,$ $\bA \bone_p = \bzero_p$. We note that for any graph with $p$ nodes, both its Laplacian matrix $\bL$ and $\{\bL^+\}^2$ are elements of $\mathcal{A}_p$. It can be seen that $\mathcal{N}(\Delta) \cap \mathcal{A}_p  = \{\bzero_{p \times p} \}$, which implies that $\forall \bA \in \mathcal{A}_p$, $\Delta(\bA) = \bzero_{p \times p} \Rightarrow \bA = \bzero_{p \times p}$. We now show that this condition implies the injectivity of $\Delta$ over $\mathcal{A}_p$.

\begin{equation}
    \begin{aligned}
    &\text{Suppose } (\forall \bA \in \mathcal{A}_p, \; \Delta(\bA) = \bzero \Rightarrow \bA = \bzero): \\
    &\text{If } \Delta(\bA_1) = \Delta(\bA_2) \text{ and } \bA_1,\bA_2 \in \mathcal{A}\\
    &\Rightarrow \Delta(\bA_1) - \Delta(\bA_2)  = \bzero \\
    &\Rightarrow \Delta(\bA_1 - \bA_2)  = \bzero \quad (\text{b.c. } \Delta \text{ is linear}) \\
    &\Rightarrow \bA_1 -\bA_2  = \bzero \quad (\text{from supposition, note that } (\bA_1 - \bA_2) \in \mathcal{A}_p)\\
    &\Rightarrow \bA_1 = \bA_2\\
    &\Rightarrow \Delta \text{ is injective over } \mathcal{A}_p
    \end{aligned}
\end{equation}

Because $\{\bL^+\}^2 \in \mathcal{A}_p$, $\{\bL^+\}^2_1 \neq \{\bL^+\}^2_2 \Rightarrow \bD_1 \neq \bD_2$. The uniqueness of the Moore-Penrose inverse in conjunction with the fact that Laplacian matrices are always positive semi-definite means that the transformation from $\bL$ to $\{\bL^+\}^2$ is injective. From the definition $\bL =\text{diag}(\bW \bone) - \bW$, it is clear that $\bW_1 \neq \bW_2 \Rightarrow \bL_1 \neq \bL_2$, and thus that $(\sigma^2_1, \bW_1) \neq  (\sigma^2_2, \bW_2) \Rightarrow \bSigma_1 \neq \bSigma_2$ for all $\sigma^2_1,\sigma^2_2 > 0$ and all $\bW_1, \bW_2 \in \mathcal{W}_G$.

\newpage
\section{MCMC Sampler}

In order to sample from the edge weight parameters of our model, we implemented the Metropolis-adjusted Langevin algorithm (MALA), which requires the analytic computation of the partial derivatives of the log posterior density with respect to all model parameters.

Below we provide details on how to obtain the partial derivative of the log-likelihood $\ell$ (assuming $\bY \sim N(\bzero, \bSigma\otimes \bI_n)$) with respect to a single edge weight $(w_{jk})$. Special thanks to Jo Eidsvik, who assisted us with computation and derivative calculations.

One may obtain the derivative $\frac{\partial\ell}{\partial w_{jk}}$ through repeated application of the chain rule as follows:

\begin{eqnarray}
\frac{\partial\ell}{\partial w_{jk}} &=& -\frac{n}{2} \mbox{trace} \left( \bSigma^{-1} \frac{\partial\bSigma}{\partial w_{jk}} \right) + \frac{1}{2} \sum_{i=1}^n \by_i' \bSigma^{-1} \frac{\partial\bSigma}{\partial w_{jk}} \bSigma^{-1} \by_i \nonumber \\
\frac{\partial\bSigma}{\partial w_{jk}} &=& -\sigma^2\frac{\partial\bD}{\partial w_{jk}} \odot \bD \odot \exp(-\bD)= -\frac{\sigma^2}{2}\frac{\partial\bD_2}{\partial w_{jk}} \odot \exp(-\bD), \nonumber  \\
\frac{\partial\bD_2}{\partial w_{jk}} &=& \mbox{diag} \left( \frac{\partial\{\bL^+\}^2}{\partial w_{jk}} \right) \mathbf{1}'_n + \mathbf{1}_n \mbox{diag} \left( \frac{\partial\{\bL^+\}^2}{\partial w_{jk}} \right)'-2\frac{\partial\{\bL^+\}^2}{\partial w_{jk}},  \\
\frac{\partial\{\bL^+\}^2}{\partial w_{jk}} &=& \bL^+ \frac{\partial\bL^+}{\partial w_{jk}} +\frac{\partial\bL^+}{\partial w_{jk}} \bL^+, \nonumber\\
\frac{d\bL^+}{\partial w_{jk}} &=& -\bL^+ \frac{\partial \bL}{\partial w_{jk}} \bL^+, \nonumber  \\
\frac{\partial \bL}{\partial w_{jk}} &=& \mbox{diag} \left( \frac{d\bW}{\partial w_{jk}} \mathbf{1}_n \right)-\frac{\partial\bW}{\partial w_{jk}} \nonumber. 
\end{eqnarray}

Here $\bD_2 = \bD \odot \bD$ and $\frac{\partial\bW}{\partial w_{jk}}$ is a $p \times p$ matrix with elements $jk$ and $kj$ equal to 1 and all others equal to 0. In practice it may be helpful to sample $\text{log}(w_{jk})$ instead to avoid boundary issues in which case $\frac{\partial\bW}{\partial\text{log}(w_{jk})}$ is a matrix with elements $jk$ and $kj$ equal to $w_{jk}$.

Assuming a Gamma$(a,b)$ prior for all edge weights, the derivative of the log-posterior with respect to $w_{jk}$ is 
\begin{equation}
   \frac{\partial\text{log} \pi(\bw|\bY)}{\partial w_{jk}} = \frac{\partial\ell}{\partial (w_{jk})} + \frac{a-1}{w_{jk}}-b.
\end{equation}

Let $\nabla_{\bw} \text{log} \pi(\bw|\bY)$ be the vector of length $e = |E|$ containing all partial derivatives of the log likelihood with respect to edge weights. The posterior for $\bw$ and $\sigma^2$ can be updated as detailed below in Algorithm 1:

\begin{algorithm}[h!]
\caption{MCMC Sampler for Covariance Model}\label{alg}
\begin{algorithmic}
\State \textbf{Input:} $\bY$ an $n \times p$ data matrix, $G$ a graph with $p$ nodes, and $s$ a tuning parameter, which may be adaptively updated during burn-in to ensure an appropriate acceptance rate.
\State \textbf{Output:} $T$ posterior samples for model parameters $\sigma^2$ and $\bw$
\State Initialize $\sigma^{2(0)}$ and $\bw^{(0)}$ and set $\bSigma^{(0)} = c_\nu(\sigma^{2(0)},\bw^{(0)})$
\For{$t = 1 \text{ to } T$}
\State Update $\sigma^{2(t)}$ by taking draw from full conditional $\pi(\sigma^{2(t)}|\bY,\bSigma^{(t-1)})$ 
\State Set $\bw^{cur} = \bw^{(t-1)}$
\State Set $\bSigma^{cur} = c_\nu(\sigma^{2(t)},\bw^{(t-1)})$
\State Propose $\bw^* \sim g(\bw^*|\bw^{cur}) = N_e(\bw^{(t-1)} + \frac{s^2}{2}\nabla_{\bw} \text{log} \pi(\bw^{cur}|\bY,\sigma^{2(t)}), s^2\bI_e)$
\State Set $\bSigma^* = c_\nu(\sigma^{2(t)},\bw^*)$
\State Compute $\alpha = \frac{ \displaystyle \pi(\bw^{cur}|\bY) g(\bw^{cur}|\bw^*) }
{ \displaystyle\pi(\bw^{cur}|\bY) g(\bw^*|\bw^{cur})  }$
\State Generate $u$ from $U(0,1)$
\If{$u < \text{min}(1,\alpha)$}
\State $\bw^{cur} \gets \bw^*$
\State $\bSigma^{cur} \gets \bSigma^*$
\EndIf
\State $\bw^{(t)} \gets \bw^{cur}$
\State $\bSigma^{(t)} \gets \bSigma^{cur}$
\EndFor
\end{algorithmic}
\end{algorithm}

\section{Code and Data}
Links to R code and data used for this article can be found at https://mfchristensen.github.io